\documentclass[aps,prd,a4paper,amsmath,showpacs,superscriptaddress,nofootinbib,preprintnumbers]{revtex4}

\pdfoutput=1
\usepackage{amssymb,amsmath,latexsym,mathrsfs}
\usepackage{graphicx,subfigure}
\usepackage{epsfig}
\usepackage{varioref,xr-hyper}
\usepackage{color}
\usepackage{multirow}
\usepackage{array}
\usepackage{hyperref}
\usepackage{wasysym}
\usepackage{color}
\usepackage{float}
\usepackage{xcolor}
\usepackage[utf8]{inputenc}
\usepackage[T1]{fontenc}

\newcommand{\be}{\begin{equation}}
\newcommand{\ee}{\end{equation}}
\begin{document}

\title{Testing Predictions of the Quantum Landscape Multiverse 3: The Hilltop Inflationary Potential}

\author{Eleonora Di Valentino}
\email{eleonora.divalentino@manchester.ac.uk}
\affiliation{ Jodrell Bank Center for Astrophysics, School of Physics and Astronomy, University of Manchester, Oxford~Road, Manchester M13 9PL, UK}

\author{Laura Mersini-Houghton}
\email{mersini@physics.unc.edu}
\affiliation{ Department of Physics and Astronomy, UNC-Chapel Hill, Chapel Hill, NC 27599, USA}

\begin{abstract}
Here we test the predictions of the theory of the origin of the universe from the landscape multiverse, against the 2015 Planck data, for the case of the Hilltop class of inflationary models, for $p=4$ and $p=6$. By considering the quantum entanglement correction of the multiverse, we can place just a lower limit on the local 'SUSY-breaking' scale, respectively $b>8.7\times10^6 GeV$ at $95 \%$ c.l. and $b>1.3\times10^8 GeV$ at $95 \%$ c.l. from Planck TT+lowP, so the case with multiverse correction is statistically indistinguishable from the case with an unmodified inflation. We find that the series of anomalies predicted by the quantum landscape multiverse for the allowed range of $b$, is consistent with Planck's tests of the anomalies. In addition, the friction between the two cosmological probes of the Hubble parameter and with the weak lensing experiments goes away for a particular subset, the $p=6$ case of Hilltop models.
\end{abstract}

\maketitle

\section{Introduction}

The final Planck likelihood for the data analysis will be released soon. So far, the existence of an intriguing series of anomalies in the CMB has been strongly evidenced by the 2015 Planck collaboration data~\cite{planck1, planck2} and recently confirmed by the new 2018 data~\cite{Akrami:2018odb,planckparams2018}. The observed anomalies are consistent with the prediction made in 2005~\cite{tomolmh1,tomolmh2,tomolmh3,tomolmh4}, and more recently in~\cite{lmh, eleonoraS,eleonoraE}, of the theory of the origin of the universe from the quantum landscape~\cite{richlmh1,richlmh2,richlmh3,archillmh1,archillmh2,archillmh3} multiverse. As we await the final Planck collaboration likelihood release, we complete our investigation of the status of the quantum landscape predictions against data, for a class of concave potential, the hilltop models~\cite{hilltop,lyth}.   

Hilltop models were investigating in detail in a series of papers in~\cite{hilltop,lyth}. They belong to the class of concave shaped inflationary models, meaning the curvature of their potential $V'' <0$. These types of potentials are favored by Planck collaboration data~\cite{planck1,Akrami:2018odb} since they produce a low tensor-to-scalar perturbations ratio $r$. Corrections to the gravitational potential in the universe, which in the theory of the quantum landscape multiverse, arise from our entanglement with all other structures in the multiverse and give rise to a series of anomalies in the CMB, were derived in~\cite{lmh,eleonoraS,eleonoraE} and analyzed for some concave inflationary models. In this paper, we derive and analyze these corrections against data in the context of Hilltop models. 

The paper is organized as follows: in the next Section we present the Hilltop model, and the corrections introduced by the entanglement in the quantum landscape multiverse to the slow-rolling inflaton field and to the inflaton potential, showing as anomalies in the CMB spectrum. In Section \ref{method} we present the method we used for analyzing the model with the data. We provide the results and the likelihood plots in Section \ref{results} for $p=4$ and Section \ref{resultsp6} for $p=6$. We conclude in Section \ref{sec:conclusions}.

\section{The Modified Hilltop Potential}
\label{sec:model}

The class of Hilltop inflationary models~\cite{hilltop,lyth} describes models where the inflaton starts rolling down from a local maximum. Some of these models are inspired by the F and D-term models of inflation and others by supergravity. The slow roll conditions for the inflaton $\phi$ are given by $\epsilon = \frac{M_p^{2}}{2}(\frac{V'(\phi)}{V(\phi)})^{2} \ll 1$, and, $\eta = M_p^2[\frac{V''(\phi)}{V(\phi)}] \ll 1$, where the unmodified potential $V(\phi)$ is of the form:
\be
V(\phi) = V_{0} [1 - \frac{1}{2}c_{hill} (\frac{\phi}{M_p})^{2}] - \lambda_{hill} (\frac{\phi^p}{M_{p}^{p-4}})  +...
\label{hilltoppotential}
\ee 
 and $M_p = 1.2209 \times 10^{19}GeV/c^2$ is the Planck mass. Here $c_{hill} = \frac{m^{2} M_{p}^{2}}{V_{0}}$, $\lambda_{hill} \ll 1$ are small parameters for slow roll to hold. In this paper, we will focus on $p = 4$ or $p=6$ models.

From the Friedman equation we have the Hubble expansion parameter
\be
3 M_{p}^2 H^2 = V(\phi)
\ee

Calculating the total number of efolds $N$ and the power spectrum $P[k]$ from here is well defined and straightforward.

In the presence of corrections, originating from quantum entanglement in the theory of the quantum landscape multiverse, which were derived in~\cite{lmh} for concave potentials~\cite{eleonoraS}, the inflaton potential $V(\phi)$ is modified to
\be \label{effectivev}
 V_{eff} = V + \frac{1}{2} \frac{V^2}{9M_{p}^4} F[b,V] = V(\phi) + f[b, V]
\ee
where $m^2 =Abs[V'']$, and, $F[b,V(\phi)]$ is the scale dependent correction term  from entanglement derived in~\cite{lmh} for concave potentials, with $b$ the landscape parameter indicating the energy scale of the local vacuum.

The energy correction term $f[b,V]$ in the effective potential, Equation (\ref{effectivev}) is

\begin{equation}
f(\phi)=\frac{1}{2} \left[ \frac{V(\phi)}{3M_P^2} \right]^2 F(\phi)
\label{energycorrection}
\end{equation}

The parameter $b$ is a landscape parameter describing SUSY-breaking scale in each vacua, therefore it varies from vacua to vacua. All the entanglement information is contained in $F[b,V]$, which was calculated from entanglement initially~\cite{tomolmh1,tomolmh2,tomolmh3,tomolmh4} and then in~\cite{lmh}. It is given by
\begin{equation}
F(\phi) = \frac{3}{2}\left(2+\frac{m^2M_P^2}{V(\phi)}\right)ln\left(\frac{b^2M_P^2}{V(\phi)}\right)-\frac{1}{2}\left(1+\frac{m^2}{b^2}\right)e^{-\frac{3b^2M_P^2}{V(\phi)}}
\end{equation}

Einstein equations get modified accordingly since the inflaton potential $V$ is now replaced by $V_{eff}$. This means the Friedmann equation for concave potentials such as the Starobinsky~\cite{eleonoraS} or the Hilltop models, becomes
\begin{equation}
3 M_{p}^2 H^2 = V_{eff} = V +f[b,V]
\label{hubble}
\end{equation}
where the correction term in the potential and its higher derivatives satisfy: $f[b,V]/V < 1$ , $df/dV < 1$, $d^2f/dV^2 < 1$, such that the slow roll condition for inflation from $V_{eff}$ holds. (This requirement places a lower bound on the parameter 'b', as shown in~\cite{tomolmh1,tomolmh2,tomolmh3,tomolmh4}.)

Power spectrum, field solution, tensor, and scalar index, now calculated from the modified potential which includes this correction term $V_{eff}$, are modified accordingly. The derivative of the effective potential is 
\begin{equation}\label{Vprime}
V'_{eff}(\phi)=V'(\phi)(1  + df(\phi)/dV)
\end{equation}
where 
\begin{equation}
\frac{df}{dV}= \frac{V}{9 M_{p}^4} \left(F[b,V] + \frac{V}{2} dF/dV \right)
\end{equation}
and
\begin{equation}
\frac{dF}{dV} = -\frac{3\left(m^2 M_{p}^2 -M_{p}^2 (b^2 +m^2) exp[-\frac{3b^2 M_{p}^2}{V}] -2V - m^2 M_{p}^2 ln[\frac{b^2 M_{p}^2}{V}]\right)}{2 V^2}\, .
\label{dFdv}
\end{equation}

The mass of the inflaton field is now obtained from the second derivative of the effective potential
\begin{equation}\label{Vsec}
V''_{eff}(\phi)=V''(\phi)\left(1 + \frac{df}{dV}\right) +\frac{d^2f}{dV^2} V'^2 \, .
\end{equation}

The unmodified field solution for the Hilltop potential is given by integrating  
\be
\frac{V}{M_{p}^2 V'} d\phi = - d ln(k)\, . 
\ee
It gives
\begin{equation}
\frac{\phi_{0}(k)}{M_{p}}= \frac{\phi_{i,0}}{M_{p}} \left(\frac{V_0}{M_{p}^{4}} \left(\frac{(k/0.002)^{2 c_{hill}}}{(x + \frac{4\lambda_{hill}}{c_{hill}}(1 - (k/0.002)^{2c_{hill}})}\right) \right)^{1/(p-2)} 
\label{purefield}
\end{equation}
where $\phi_{i,0}$ is the initial field value at the onset of slow roll. We take it here to be $\phi_{i,0}= \frac{M_P}{\sqrt{c_{hill}}}$ and $x=\frac{12\lambda_{hill}}{2c_{hill}(1-c_{hill})}$. 

In the presence of modifications, the potential is replaced by $V_{eff}$, therefore the modified field solution obtained by 
\begin{equation}
3 H d\phi /dt = -\frac{\partial V_{eff}}{\partial d\phi}
\label{fieldsol}
\end{equation}
satisfies the equation
\begin{equation} \label{fieldwithk}
\frac{V_{eff}}{M_{p}^{2} V'_{eff}} d\phi = - dln(k)
\end{equation}
which we integrate to obtain the modified field solution.

Again, in the latter we used $d\phi/dt = H d\phi/dlnk$ to have a $k-$dependent field Equation (\ref{fieldsol}). Please note that the correction term originating from quantum entanglement contained in the complicated expression for $F[b,V(\phi)]$, or accordingly $f[b,V(\phi)]$, is nonlocal and $k-$dependent. Since it is a derived quantity, not a phenomenological one, its expression leaves no room for tweaking or changing it. Therefore, the series of anomalies induces in the CMB spectrum,  we discuss in the next sections below, originating from this term, are also scale dependent and robust predictions (meaning we cannot change them to fit the data).

Since we require that slow roll holds even with correction terms, then we can approximate the integral in Equation (\ref{fieldwithk}) as: 
\be
\int \frac{V_{eff}}{V'_{eff}} d\phi \simeq \left(\frac{1+ f/V}{1+ df/dV}\right)\int \frac{V}{V'}\,.
\ee 

The Equation (\ref{fieldwithk}) gives us the field as a function of $k$, or equivalently the number of efolds $dN$, since it allows us to also integrate $dN$ from start to end of slow roll to get the number of efolds $N_{star}$.

For the Hilltop potential of Equation (\ref{hilltoppotential}), following the above derivation, results in this modified inflaton field solution 
\begin{equation}
 \frac{\phi(k)}{M_{p}} = \frac{{\tilde \phi_{i}}}{M_{p}} \left(\frac{V_0}{M_{p}^{4}} \left( \frac{(k/0.002)^{2 {\tilde c_{hill}}}}{({\tilde x} + \frac{4 \lambda_{hill}}{{\tilde c_{hill}}}(1 - (k/0.002)^{2{\tilde c_{hill}}})}\right)\right)^{1/(p-2)} 
\label{modifiedfield}
\end{equation}

Here the fiducial mode is $k_{\*} =0.002$ and the ``tilde'' quantities, are modified quantities of their corresponding unmodified quantities. For example, we have:

\be
{\tilde c_{hill}} = c_{hill} \frac{(1 + df/dV)}{(1 + \frac{f}{V})}\,, 
\ee
\be
\tilde{\eta}  =M_{p}^{2} \frac{V''_{eff}}{V_{eff}}\,, 
\ee
\be
\tilde{\epsilon} = \frac{M_{p}^{2}}{2} \frac{V'_{eff}}{V_{eff}} \,, 
\ee
\be
{\tilde \phi_{i}} = (\frac{1}{c_{hill}}\frac{1+f/V}{1+ df/dV})^{1/(p-2)}\,, 
\ee
\be
{\tilde x} = \frac{12\lambda_{hill}}{2{\tilde c_{hill}}(1-{\tilde c_{hill}})}\,.
\ee 
where $V_{eff}$, $V'_{eff}$ and $V''_{eff}$ are given respectively from Equations~(\ref{effectivev}),~(\ref{Vprime}) and (\ref{Vsec}).
Please note that the correction to the field solution contains $(1+df/dV)/(1+f/V)$. Here we demand that $\tilde{c}_{hill},\, \tilde\epsilon,\, \tilde\eta$ continue to satisfy the slow roll conditions, which, as mentioned, allows us to find a lower bound on the parameter $b$ which, in combination with the inflaton potential V, controls the strength of the corrections in $V_{eff}$. 

We can now put everything together to calculate the power spectrum and tensor-to-scalar ratio. The expression below for the power spectrum uses the reduced Planck mass $M=2.435 \times 10^{18} GeV/c^2$ instead of Planck mass $M_{p}$, thus the change in notation from $M_p$ to $M$ (In our analysis all the factors of $2 \pi$ are carefully taken into account going from $M$ to $M_p$ to ensure consistency.). The modified power spectrum $P_{\zeta}(k)$, related to its unmodified spectrum $P_{0}(k)$ is
\begin{equation}
P_\zeta(k)=\frac{1}{24\pi^2M_{p}^6}\left[ \frac{V_{eff}(\phi)^3}{V'_{eff}(\phi)^2} \right] \simeq P_{0}[k] \frac{(1 + f/V)^3}{(1 + df/dV)^2}
\end{equation}
In our notation, unmodified fields and spectra are denoted by $_0$, e.g., $\phi_0 , r_0 , P_0$ and are evaluated with respect to unmodified field. All modified quantities $\phi, r, P$ etc., denoted without the $_0$, are evaluated with respect to the modified field $\phi$, and not $\phi_0$.

The modified tensor-to-scalar ratio is 
\begin{equation} 
r [k] = 8 M_{p}^2 \left(\frac{V'_{eff}(\phi)}{V_{eff}(\phi)}\right)^{2} 
\end{equation}

Using $V'$ and $V'_{eff}$ which we calculated above, we have approximately:

\be
r[k] = r_{0}[k]  (\frac{( 1+ df/dV )}{( 1 + f/V )})^2\,.
\ee

We can also calculate $n[k] - 1 = dln(P[k])/dln(k)$ from the expression for $P[k]$ above. Then the unmodified scalar tensor is $n_{0}[k] -1 =  dln(P_0[k])/dln(k)$ where the $_0$ notation means unmodified field and power spectrum and scalar tensor.

With these expressions, we are now ready to check the status of the {\it modified} Hilltop models against data, for the cases $p=4$ and $p=6$: in order to scrutinize the predictions of anomalies from the quantum landscape multiverse; and, check whether the allowed range of the landscape vacuum energy from which our universe inflated, given by the parameter $b$ that controls the corrections in $V_{eff}$, still satisfies the slow roll conditions.

\section{Analysis Method} \label{method}
We will explore the modified Hilltop model by considering the $4$ standard cosmological parameters and $4$ inflationary parameters. These are, respectively, the baryon  energy density $\Omega_bh^2$; the cold dark matter energy density $\Omega_ch^2$; the reionization optical depth $\tau$; the ratio between the sound horizon and the angular diameter distance at decoupling $\Theta_{s}$; the logarithm of the SUSY-breaking scale associated with the landscape effects $log(b[GeV])$; the energy scale of the inflation $10^{12}V_0/M^4$, $10^{11}\lambda_{hill}$ and $c_{hill}$.

Furthermore, we will also consider a couple of extensions to this baseline model, by adding one more parameter per time, namely the effective number of relativistic degrees of freedom $N_{\rm eff}$ and the dark energy equation of state $w$. 
All the parameters are explored within the range of the conservative flat priors reported in Table~\ref{priors}, for $p=4$ and $p=6$.

\begin{figure}[]
\centering
\includegraphics[width=7.0cm]{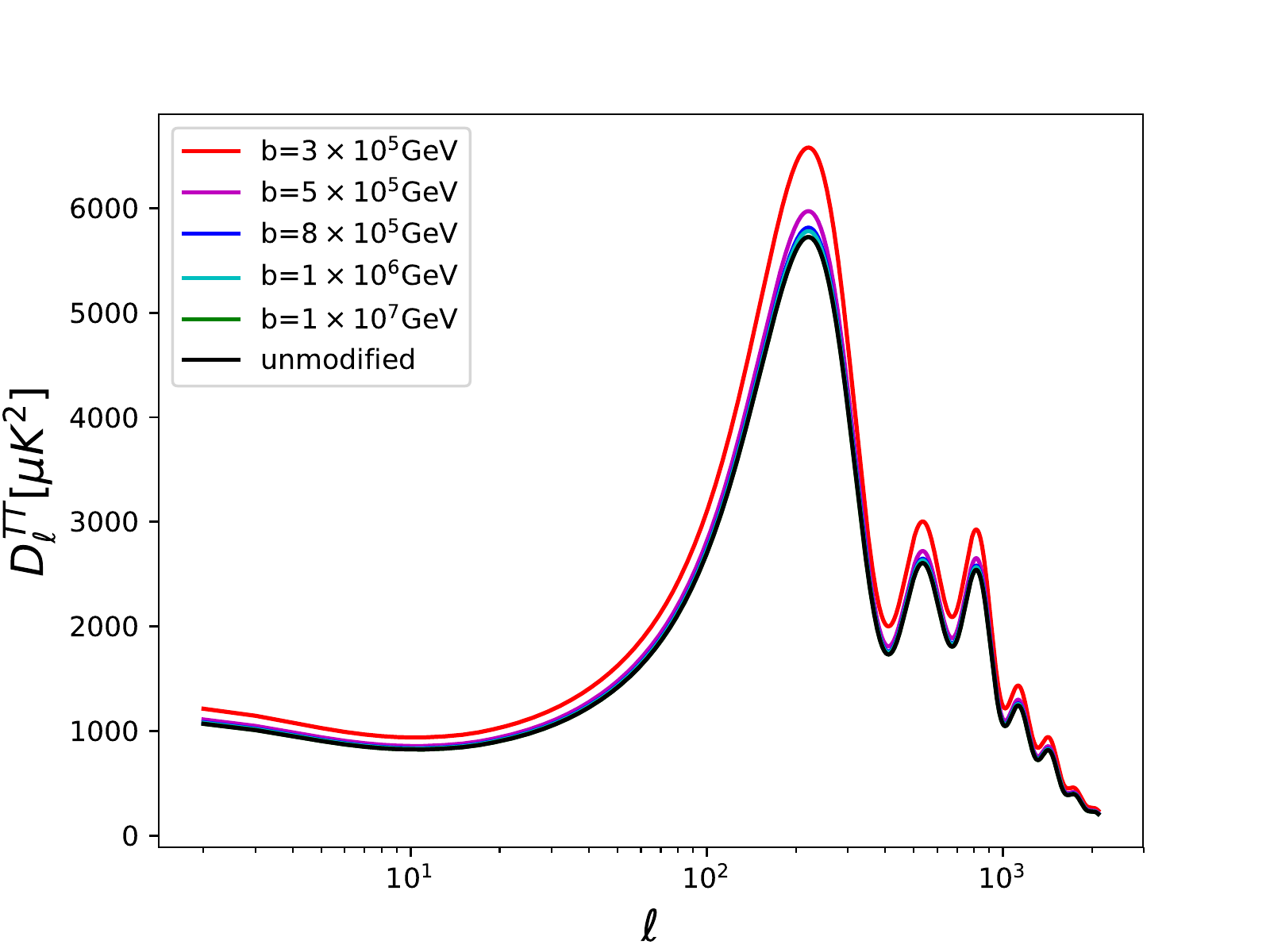}
\includegraphics[width=7.0cm]{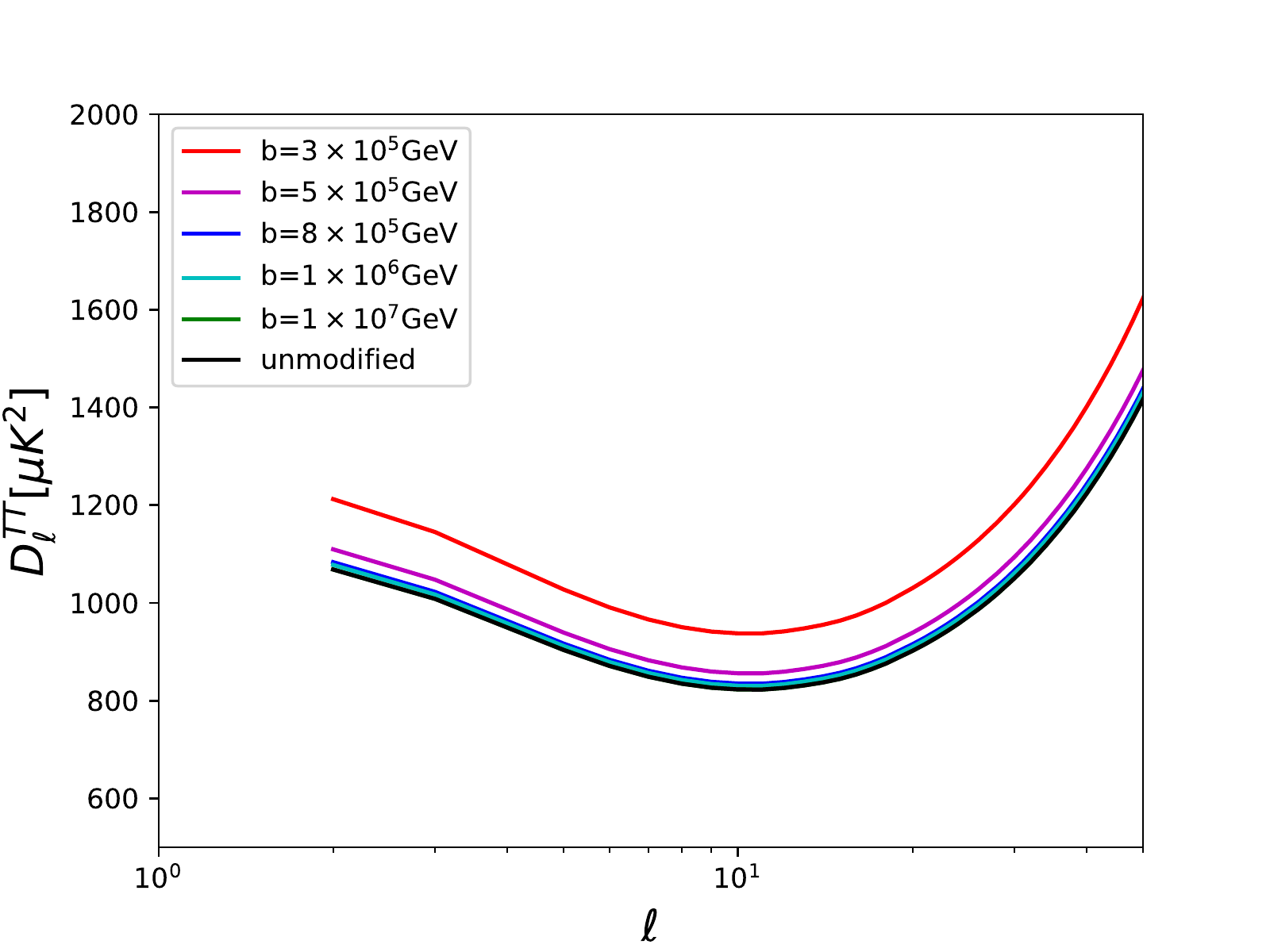}
\includegraphics[width=7.0cm]{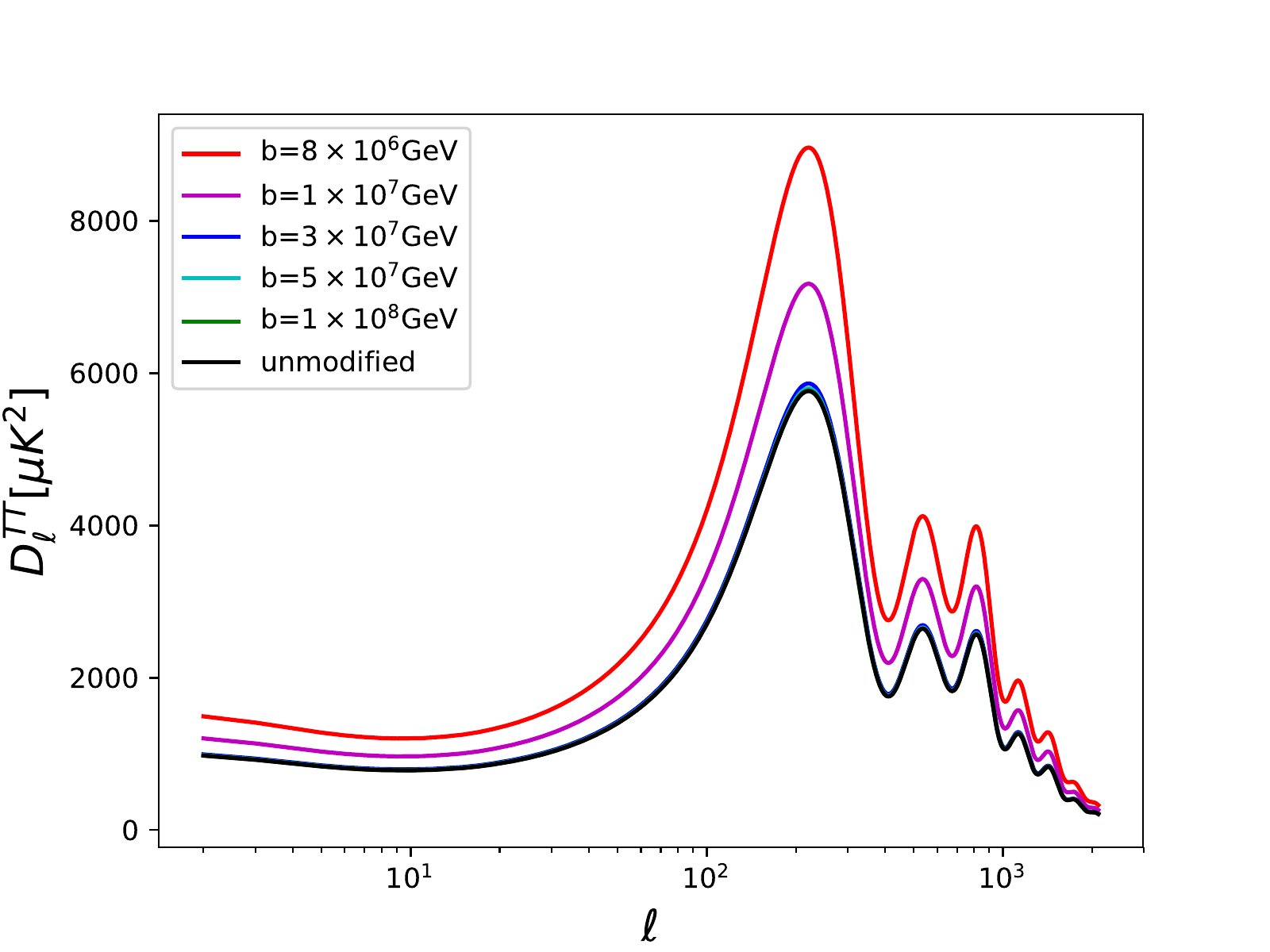}
\includegraphics[width=7.0cm]{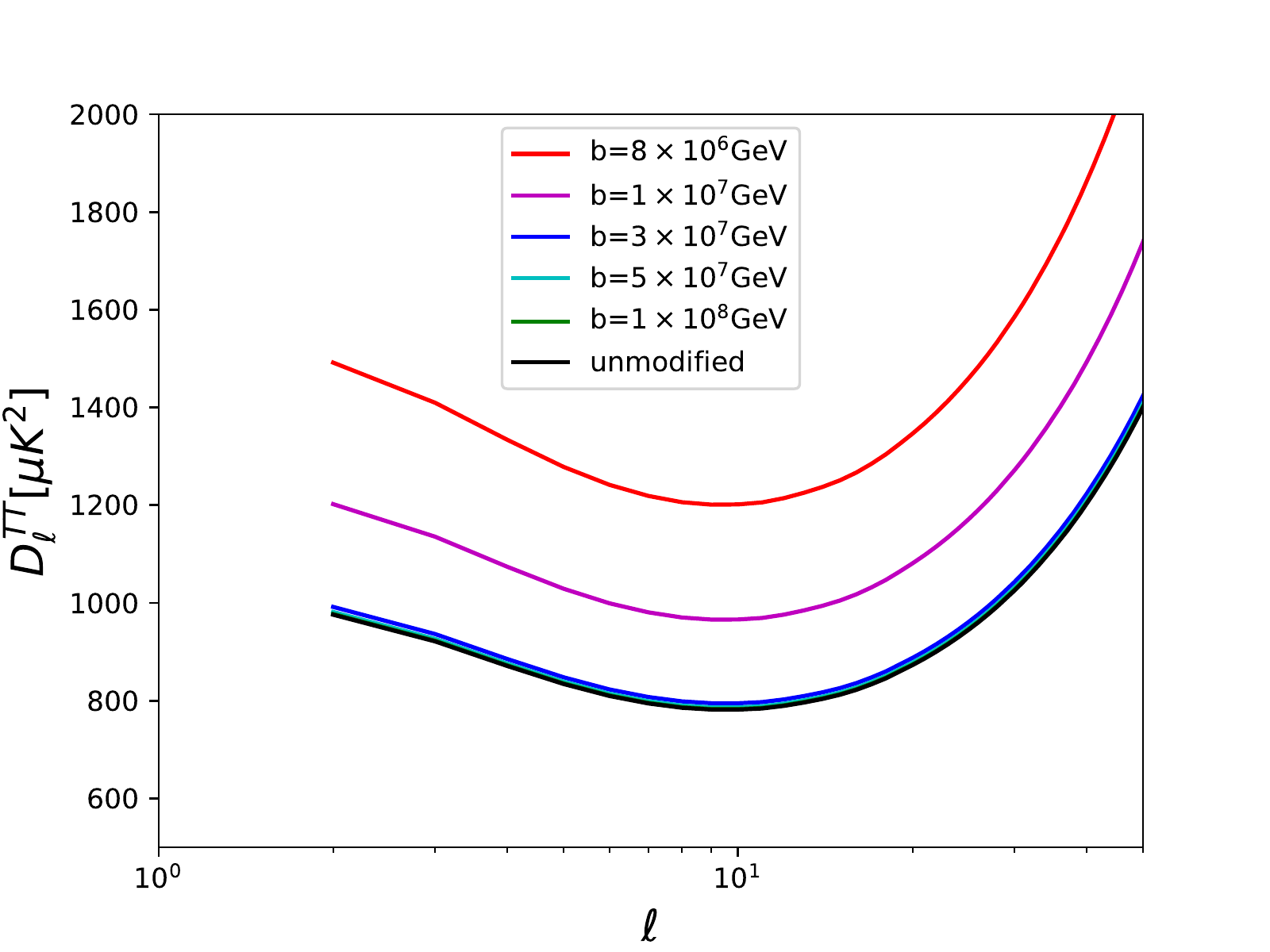}
\caption{The temperature CMB angular power spectrum by varying the SUSY-breaking scale $b$ associated with the landscape effects, for the modified Hilltop model with $p=4$ ({\bf upper panels}) and $p=6$ ({\bf bottom panels}).}
\label{clparamST}
\end{figure}

\begin{figure}[]
\centering
\includegraphics[width=7.5cm]{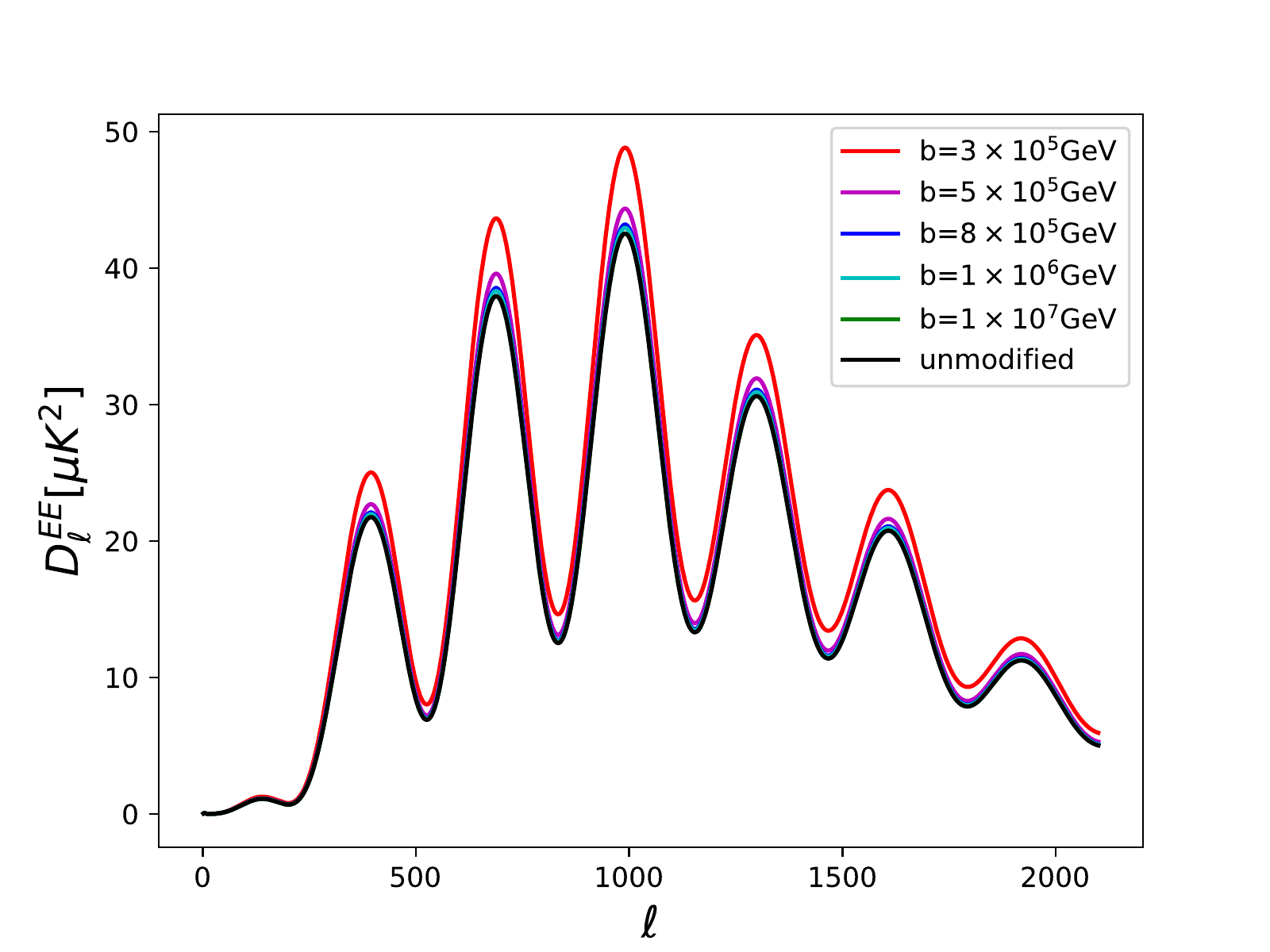}
\includegraphics[width=7.5cm]{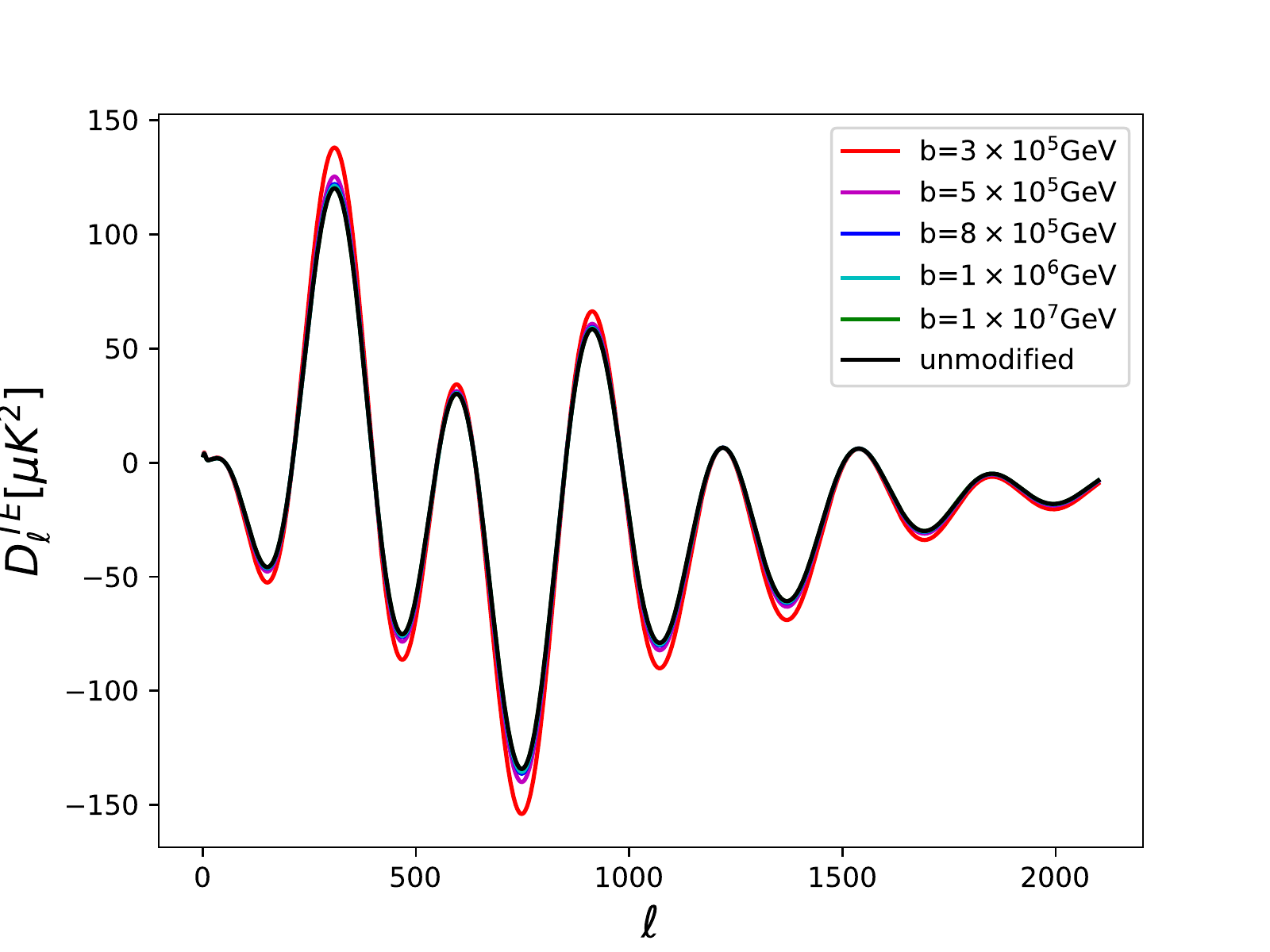}
\includegraphics[width=7.5cm]{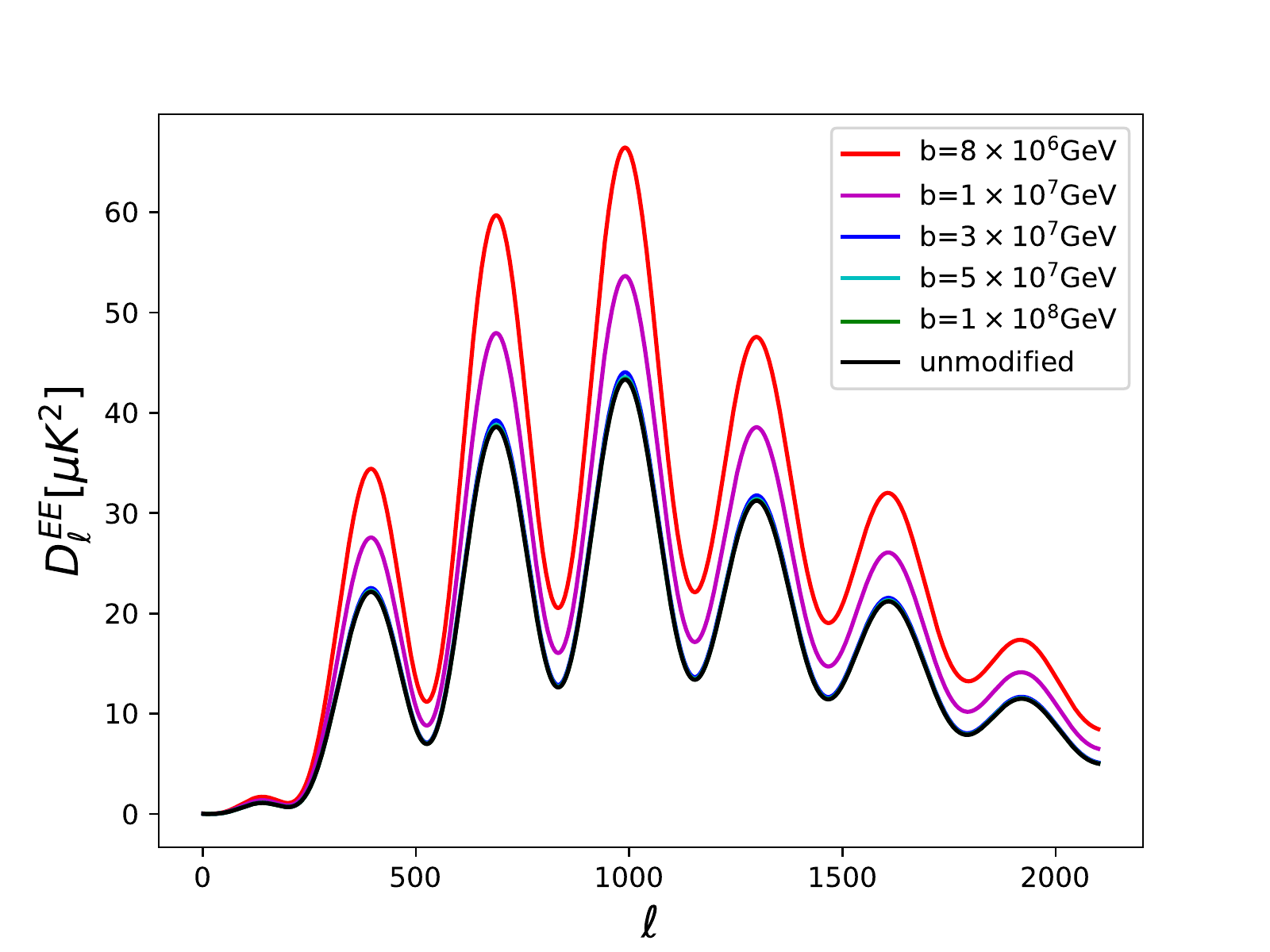}
\includegraphics[width=7.5cm]{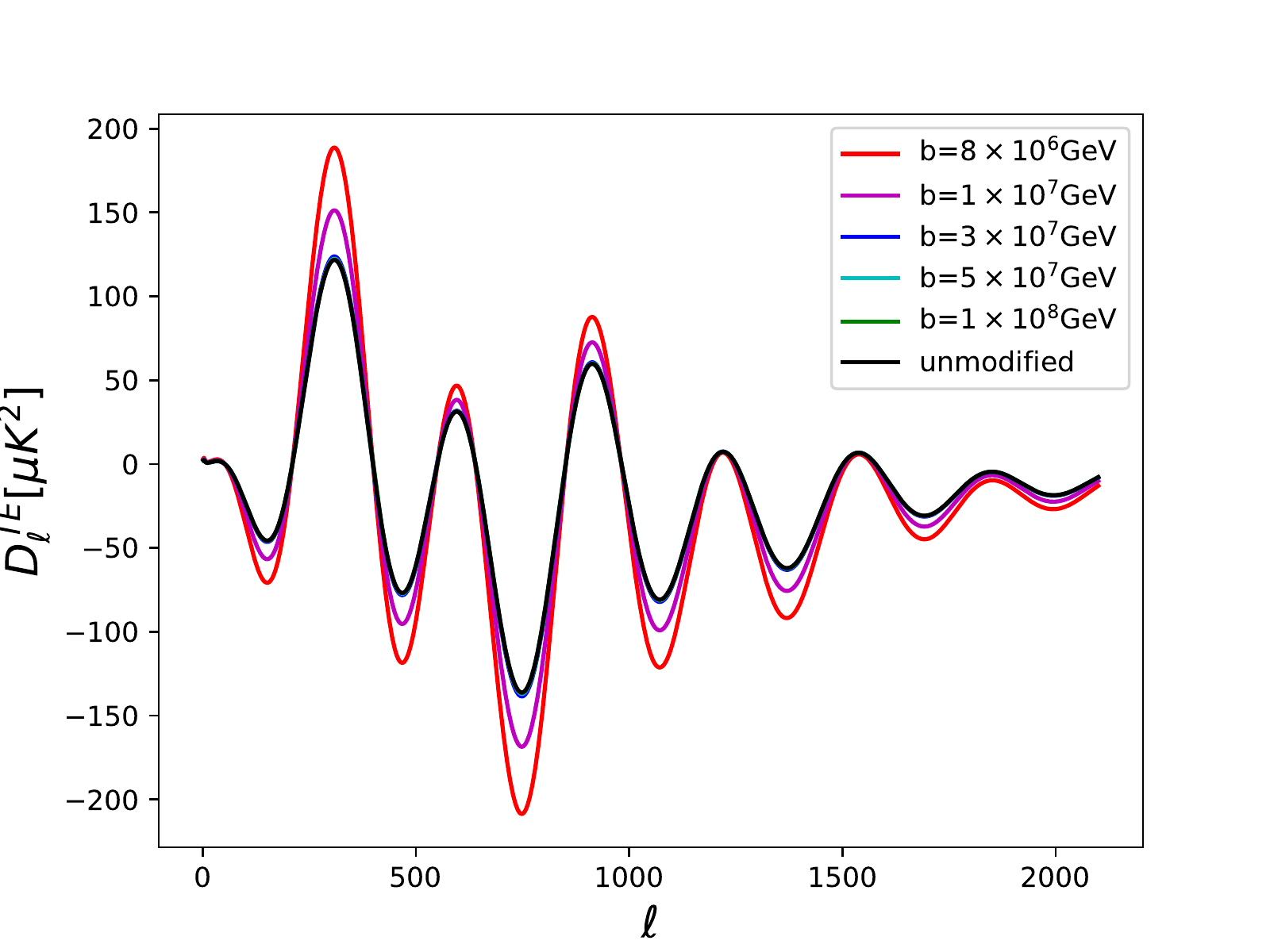}
\caption{The polarization CMB angular power spectra by varying the SUSY-breaking scale $b$ associated with the landscape effects, for the modified Hilltop model with $p=4$ ({\bf upper panels}) and $p=6$ ({\bf bottom~panels}).}
\label{clparamSP}
\end{figure}

\begin{figure}[]
\centering
\includegraphics[width=7.5cm]{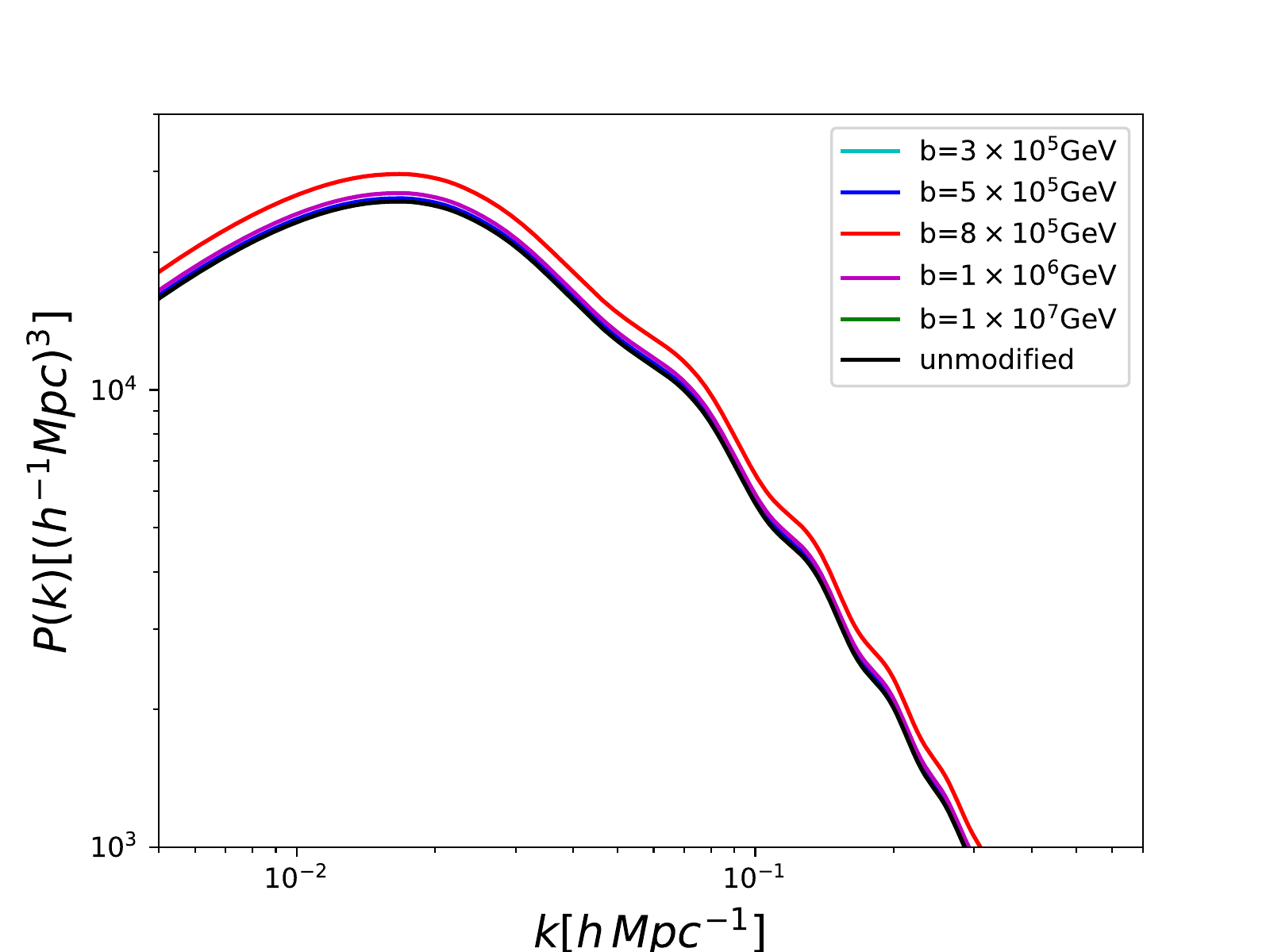}
\includegraphics[width=7.5cm]{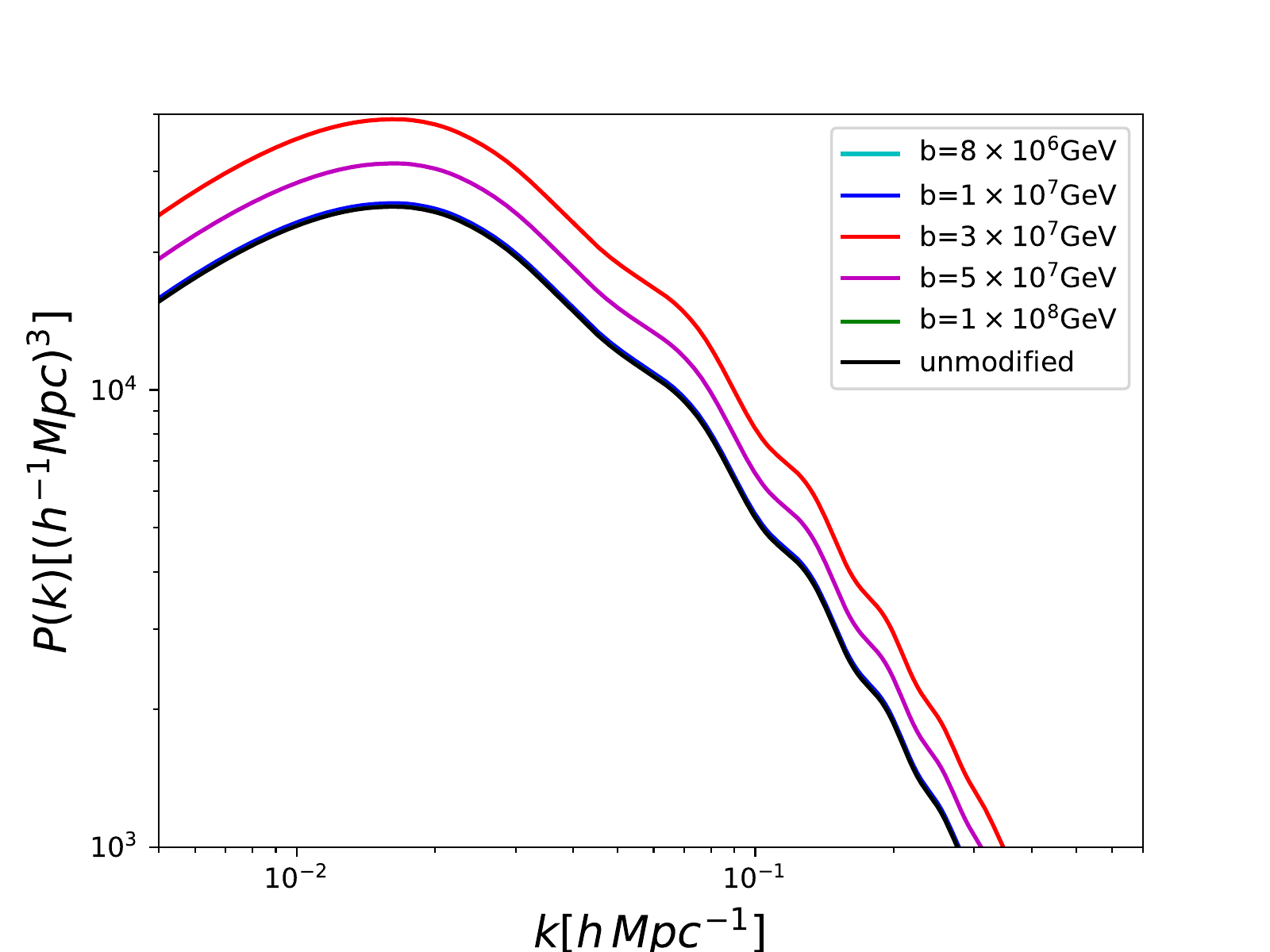}
\caption{The matter power spectrum by varying the SUSY-breaking scale $b$ associated with the landscape effects, for the modified Hilltop model with $p=4$ ({\bf left panel}) and $p=6$ ({\bf right~panel}).}
\label{clparamSM}
\end{figure}

We show the manner in which the SUSY-breaking scale $b$ affects the CMB temperature, polarization, and matter power spectra, respectively in Figures~\ref{clparamST}--\ref{clparamSM}: the effect on the power spectra, when increasing the value of $b$, is an overall decreasing of all the peaks.
\begin{table}[]
\centering
\caption{External priors on the cosmological parameters assumed in this work.}
\label{priors}
\begin{tabular}{c|c|c}
\hline
{\bf Parameter}                    & \boldmath{$p=4$} & \boldmath{$p=6$} \\
\hline
$\Omega_{\rm b} h^2$          & $[0.005,0.1]$ & $[0.005,0.1]$\\
$\Omega_{\rm cdm} h^2$       & $[0.001,0.99]$ & $[0.001,0.99]$\\
$\Theta_{\rm s}$             & $[0.5,10]$ & $[0.5,10]$\\
$\tau$                       & $[0.01,0.8]$ & $[0.01,0.8]$\\
$log(b[GeV])$                        & $[1, 19]$ & $[1, 19]$\\
$10^{12}V_0/M^4$         & $[0.02,40]$ & $[10,80]$\\
$10^{11}\lambda_{hill}$         & $[0.2,0.6]$ & $[0.2,1.0]$\\
$c_{hill}$         & $[0,1]$ & $[0,0.05]$\\
$N_{\rm eff}$ & $[0.05,10]$ & $[0.05,10]$\\
$w$ & $[-3.0,0.3]$ & $[-3.0,0.3]$\\
\hline
\end{tabular}

\end{table}
\unskip

We constrained the parameters listed before by considering the following cosmological probes.
Firstly, we analyzed the ``Planck TT + lowP'' data, i.e., the full range of the 2015 temperature power spectrum ($2\leq\ell\leq2500$) combined with the low-$\ell$ polarization power spectra in the multipoles range $2\leq\ell\leq29$ provided by the Planck collaboration~\cite{Aghanim:2015xee}.
Secondly, we included the high multipoles Planck polarization data~\cite{Aghanim:2015xee}, in the range $30\leq\ell\leq2500$, and we called this combination ``Planck~TTTEEE + lowP''. 
Then, we replaced the low-$\ell$ data in the multipoles range $2\leq\ell\leq29$ with a gaussian prior on the reionization optical depth $\tau=0.055\pm0.009$, as obtained from Planck HFI measurements~\cite{newtau}, and we called this prior ``tau055''.
Finally, we added the baryon acoustic oscillation data from 6dFGS~\cite{beutler2011}, SDSS-MGS~\cite{ross2014}, BOSSLOWZ~\cite{anderson2014} and CMASS-DR11~\cite{anderson2014} surveys as was done in~\cite{planckparams2015}, and we referred to this dataset as ``BAO''.

To analyze statistically these data exploring the modified Hilltop model for the entanglement, we have used the publicly available Monte-Carlo Markov Chain package \texttt{cosmomc}~\cite{Lewis:2002ah}, with a convergence diagnostic based on the Gelman and Rubin statistic, where we modified the CAMB code~\cite{Lewis:1999bs}, to include the primordial power spectrum of our model. It implements an efficient sampling of the posterior distribution using the fast/slow parameter decorrelations~\cite{Lewis:2013hha}, and it includes the support for the Planck data release 2015 Likelihood Code~\cite{Aghanim:2015xee} (see \url{http://cosmologist.info/cosmomc/}). 

\section{Results $p=4$}\label{results}
The result of all the explorations are given in Tables~\ref{tablerp4}--\ref{tablewp4}, where we report the constraints at $95 \% $ c.l. on the cosmological parameters. All the bounds that we will quote hereinafter there will be at $95\%$ c.l., unless otherwise expressed. These tables differ for the cosmological scenario explored, respectively the $\Lambda$CDM+r, $\Lambda$CDM+r+$N_{\rm eff}$, $w$CDM+r. In Table~\ref{tableunmod_p4} we can see instead the bounds for the same cases for the unmodified Hilltop scenario with $p=4$.

\begin{table}[]
\caption{$95 \% $ c.l. constraints on cosmological parameters in our baseline $\Lambda$CDM+r scenario from different combinations of datasets with a modified Hilltop inflation with $p=4$.}
\label{tablerp4}
\centering
\scalebox{0.89}{\begin{tabular}{lccccccc}
\hline
      {\bf Planck TT}    & &{\bf Best Fit}&&{\bf Best Fit}&&{\bf Best Fit} \\                     
 & {\bf + lowP}   &{\bf + lowP} &        {\bf + lowP + BAO} &{\bf + lowP + BAO}&{\bf + tau055}&{\bf + tau055} \\

$\Omega_{\textrm{b}}h^2$& $0.02225^{+0.00047}_{-0.00044} $&$0.02225$& $0.02227\,^{+0.00040}{-0.00039}$  &$0.02224$  & $0.02210^{+0.00042}_{-0.00044} $ &$0.02193$  \\

$\Omega_{\textrm{c}}h^2$& $0.1195^{+0.0044}_{-0.0042}$&$0.1199$& $0.1189\,^{+0.0025}_{-0.0026}$ &$0.1195$   & $0.1213^{+0.0044}_{-0.0042} $ &$0.1228$  \\

$\tau$& $0.078^{+0.037}_{-0.037}$&$0.076$& $0.080\,^{+0.036}_{-0.034}$ &$0.084$   & $0.059\,^{+ 0.017}_{-0.018}$  &$0.059$ \\

$10^{12}V_0/M^4$& $<11.7$&$0.70$& $<12.5$ &$0.53$   & $<29.2$&$5.6$   \\

$log(b[GeV])$& $>6.75$&$11.8$& $>6.86$   &$13.3$ & $>7.44$ &$10.2$  \\

$10^{11}\lambda_{hill}$& $0.304^{+0.059}_{-0.048}$&$0.276$& $0.304^{+0.063}_{-0.048}$ &$0.280$   & $0.36^{+0.12}_{-0.10}$ &$0.31$  \\

$c_{hill}$& $0.0031\,\pm0.0012$&$0.0033$& $0.00295^{+0.00094}_{-0.00089}$  &$0.0034$  & $0.0033^{+0.0013}_{-0.0012}$ &$0.0037$  \\

$r$ &  $<0.0941$ &$0.0061$&  $<0.101$&$0.0046$  & $<0.219$&$0.047$\\

$H_0$ &      $67.4 \pm 1.9$&$67.2$&      $ 67.7^{+1.2}_{-1.1}$ &$67.5$& $ 66.6^{+1.8}_{-1.8}$  &$66.0$   \\

$\sigma_8$   & $ 0.829 ^{+0.029}_{-0.028}$  &$0.830$ & $ 0.828\,^{+0.029}_{-0.028}$ &$0.836$  & $ 0.820^{+0.021}_{-0.020}$&$0.826$  \\

$\chi^2$   && $ 11266.6$   && $ 11271.1$   && $ 771.4$  \\

\hline

     {\bf Planck TTTEEE}    & & {\bf Best Fit} &&  {\bf Best Fit} &&  {\bf Best Fit} \\                     
 &  {\bf + lowP}   &  {\bf + lowP}  &        {\bf + lowP + BAO } &  {\bf + lowP + BAO} &  {\bf + tau055} &  {\bf + tau055} \\

$\Omega_{\textrm{b}}h^2$& $0.02224^{+0.00032}_{-0.00032} $&$0.02227$& $0.02228^{+0.00028}_{-0.00026} $  &$0.02234$  & $0.02216^{+0.00030}_{-0.00028} $ &$0.02226$  \\

$\Omega_{\textrm{c}}h^2$& $0.1198^{+0.0028}_{-0.0030}$&$0.1193$& $0.1192^{+0.0021}_{-0.0021}$   &$0.1181$ & $0.1208\,\pm0.0027 $  &$0.1205$ \\

$\tau$& $0.078 ^{+0.034}_{-0.033}$&$0.091$& $0.082^{+0.032}_{-0.032}$  &$0.091$  & $0.061^{+0.017}_{-0.016}$ &$0.073$  \\

$10^{12}V_0/M^4$& $<13.4$&$0.18$& $<12.4$  &$0.17$  & $<26.6$  &$11.6$ \\

$log(b[GeV])$& $>6.90$&$15.7$& $>6.80$  &$14.8$  & $>7.24$ &$16.8$  \\

$10^{11}\lambda_{hill}$& $0.310^{+0.068}_{-0.049}$&$0.279$& $0.307^{+0.063}_{-0.047}$  &$0.274$  & $0.35^{+0.11}_{-0.09}$ &$0.362$  \\

$c_{hill}$& $0.00318^{+0.00094}_{-0.00099}$&$0.0033$& $0.00306^{+0.00085}_{-0.00079}$  &$0.0030$  & $0.0033\pm0.0011$ &$0.0033$  \\

$r$ &  $<0.106$ &$0.0016$&  $<0.0983$ &$0.0014$ & $<0.199$&$0.091$\\

$H_0$ &      $67.3^{+1.4}_{-1.2}$&$67.4$&      $ 67.54^{+0.93}_{-0.94}$&$68.0$ & $ 66.8\,\pm1.2$  &$67.1$   \\

$\sigma_8$   & $ 0.830 ^{+0.026}{-0.025}$ &$0.839$  & $ 0.831\,\pm0.026$  &$0.833$ & $ 0.820\,^{+0.016}_{-0.015}$ &$0.827$ \\

$\chi^2$   && $ 12943.7$  & & $ 12949.1$ &  & $ 2451.7$  \\

\hline

\end{tabular}}

\end{table}

If we compare Table~\ref{tablerp4}, where there are the constraints for this modified Hilltop inflation with $p=4$, and the first 2 columns of Table~\ref{tablelcdm}, where they are the constraints in the standard $\Lambda$CDM+r scenario, we have very robust constraints for all the cosmological parameters with no significant departure from their values with respect to the standard case. Moreover, these bounds are also perfectly consistent with the same cases for the original Hilltop model with $p=4$, how can be seen by looking at Table~\ref{tableunmod_p4}. However, for our modified Hilltop inflation the $\chi^2$ gets worse, even if with more degrees of freedom.

\begin{figure}[]
\centering
\includegraphics[scale=0.6]{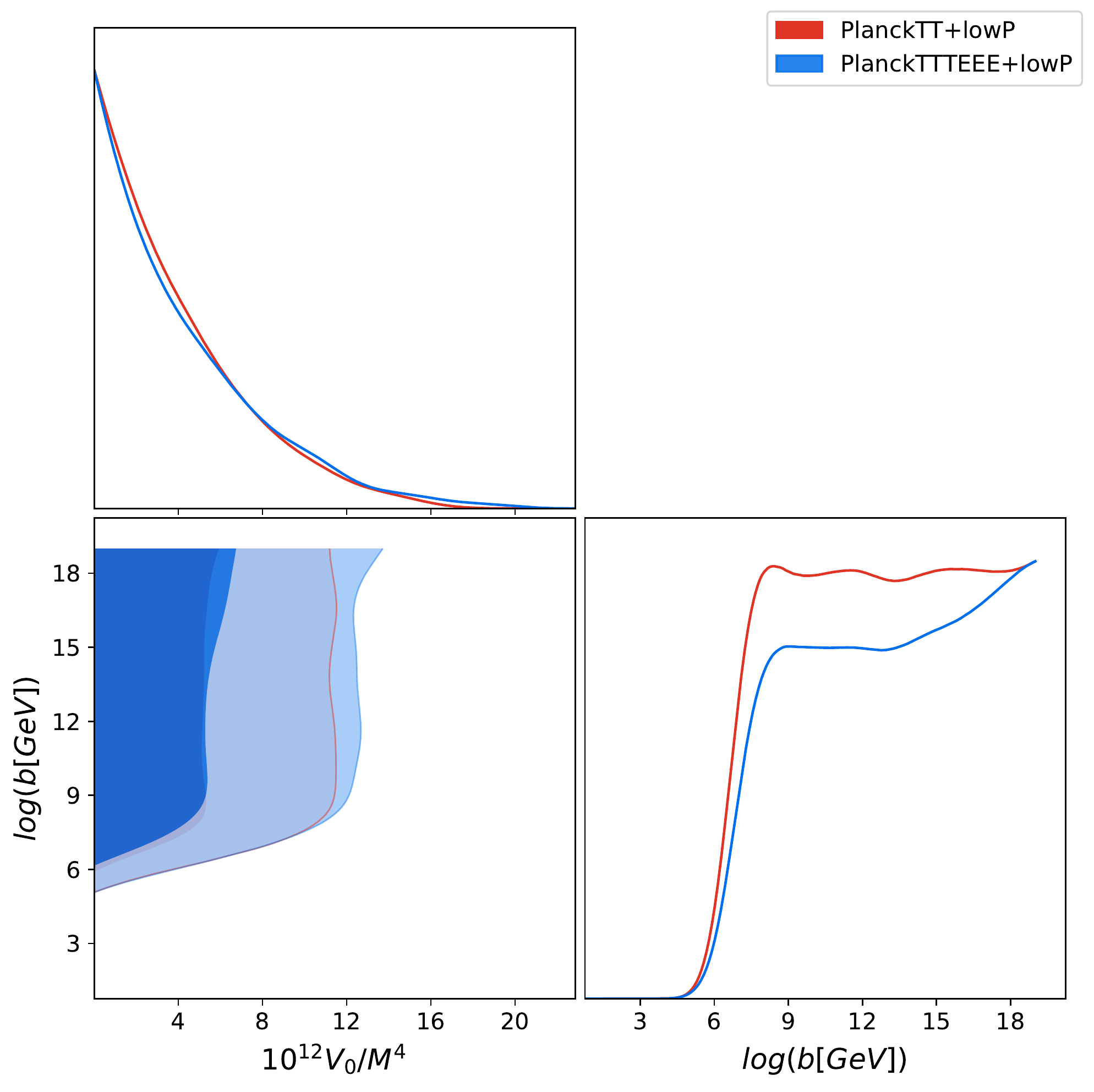}
\caption{Constraints at $68 \%$ and  $95 \%$ confidence levels on the $12^{10}V_0/M^4$ vs. $log(b[GeV])$ plane, in our modified $\Lambda$CDM+r Hilltop inflation with $p=4$. Looking at the Table~\ref{tablerp4}, we can see that the best fits for these parameters are $12^{10}V_0/M^4 = 0.70$ and $log(b[GeV])=11.8$ for PlanckTT+lowP, while they are $12^{10}V_0/M^4 = 0.18$ and $log(b[GeV])=15.7$ for PlanckTTTEEE+lowP.}
\label{figv0bp4}
\end{figure}

Regarding the inflationary parameters that describe the theory analyzed here, we have an upper limit of the tensor-to-scalar ratio $r$, consistent with the $\Lambda$CDM+r value. We find for this model and Planck TT + lowP that $r<0.0941$ c.l.. If we look at Figure~\ref{figv0bp4}, which shows the constraints at $68 \%$ and  $95 \%$ confidence levels on the $10^{12}V_0/M^4$ vs. $log(b)$ plane, we can see that there exists a lower limit for $b>5.6\times10^6 GeV$ and $V_0<11.7\times 10^{-12} M_P^4$ for Planck TT+lowP.

\begin{figure}[]
\centering
\includegraphics[scale=0.5]{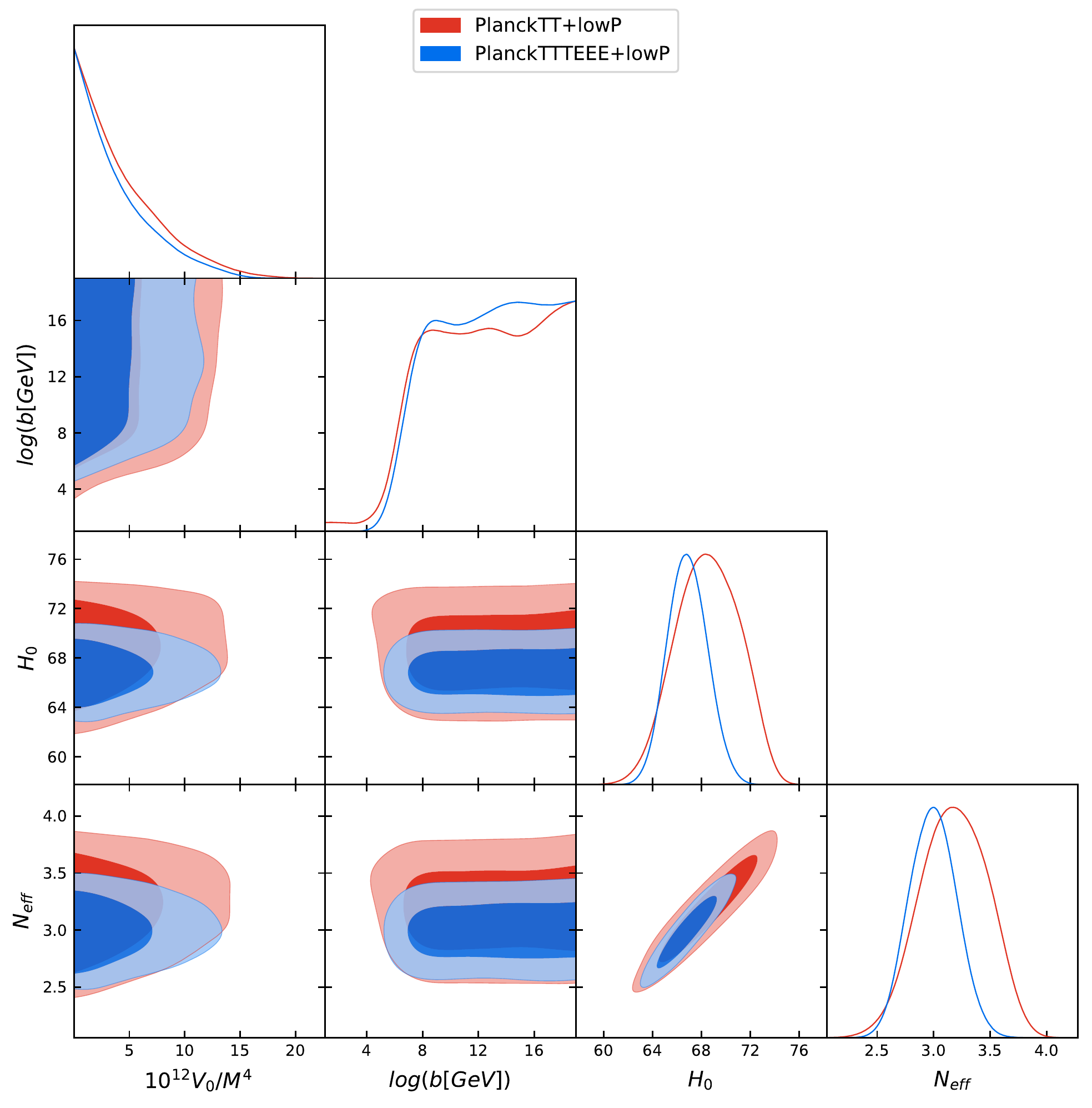}
\caption{Constraints at $68 \%$ and  $95 \%$ confidence levels in our modified $\Lambda$CDM+r+$N_{\rm eff}$ Hilltop scenario with $p=4$. Looking at the Table~\ref{tablennup4}, we can see that the best fits for these parameters are $12^{10}V_0/M^4 = 0.49$, $log(b[GeV])=15.3$, $N_{\rm eff}=3.00$ and $H_0=66.9$ for PlanckTT+lowP, while they are $12^{10}V_0/M^4 = 0.25$, $log(b[GeV])=8.4$, $N_{\rm eff}=3.01$ and $H_0=66.9$ for PlanckTTTEEE+lowP.}
\label{fignnup4}
\end{figure}

Moreover, by introducing a dark radiation component free to vary $N_{\rm eff}$ in this modified Hilltop scenario with $p=4$,  we have very robust constraints for all the cosmological parameters, see Table~\ref{tablennup4}, which have no significant shifts with respect to the standard $\Lambda$CDM+$N_{\rm eff}$ model (see the columns 3 and 4 of Table~\ref{tablelcdm}), and with respect to the original Hilltop model with $p=4$ (see Table~\ref{tableunmod_p4}). However, also in this case, for our modified Hilltop inflation the $\chi^2$ gets worse. The reason we are introducing this extra parameter is that in the minimal standard cosmological model or in other inflationary models (for example~\cite{Tram:2016rcw,DiValentino:2016ucb}), to let $N_{\rm eff}$ free to vary produces a value for this neutrino effective number higher than its expected value $3.045$~\cite{Mangano:2005cc,deSalas:2016ztq}, and the shift of the parameters correlated~\cite{darkradiation,edv1,ma1,DiValentino:2016ucb,DiValentino:2017oaw}. In particular, it is interesting to note the shift towards higher values of the Hubble constant $H_0$ (see Figure~\ref{fignnup4}) that could help in solving the tension now at $3.8\sigma$ between the constraints coming from the Planck satellite~\cite{planckparams2013,planckparams2015,planckparams2018} and the local measurements of the Hubble constant of Riess et al.~\cite{R11,R16,R18}. In the modified Hilltop scenario, we find the same effect when just the PlanckTT+lowP is considered, and the tension on the Hubble constant becomes of $1.6\sigma$. When introducing the polarization data, the tension with Planck is restored at $2.6\sigma$.

\begin{table}[]
\caption{$95 \% $ c.l. constraints on cosmological parameters in our baseline $\Lambda$CDM+r+$N_{\rm eff}$ scenario from different combinations of datasets with a modified Hilltop inflation  with $p=4$.}
\label{tablennup4}
\centering
\scalebox{0.78}{\begin{tabular}{lccccccc}
\hline
      {\bf Planck TT}    & &{\bf Best Fit}&&{\bf Best Fit}&&{\bf Best Fit} \\                     
 & {\bf + lowP}   &{\bf + lowP} &        {\bf + lowP + BAO} &{\bf + lowP + BAO}&{\bf + tau055}&{\bf + tau055} \\

$\Omega_{\textrm{b}}h^2$& $0.02236^{+0.00065}_{-0.00067} $&$0.02204$& $0.02236^{+0.00048}_{-0.00047}$  &$0.02221$  & $0.02183^{+0.00069}_{-0.00072} $  &$0.2154$ \\

$\Omega_{\textrm{c}}h^2$& $0.1210^{+0.0075}_{-0.0077}$&$0.1203$& $0.1210^{+0.0078}_{-0.0075}$    &$0.1191$& $0.1178^{+0.0077}_{-0.0078} $   &$0.1151$\\

$\tau$& $0.082^{+0.041}_{-0.042}$&$0.078$& $0.082^{+0.035}_{-0.036}$   &$0.075$ & $0.058\,^{+0.017}_{-0.018}$ &$0.062$  \\

$10^{12}V_0/M^4$& $<11.7$&$0.49$& $<11.7$   &$1.8$ & $<27.1$  &$7.3$ \\

$log(b[GeV])$& $>6.42$&$15.3$& $>6.42$   &$18.2$ & $>7.11$  &$11.7$ \\

$10^{11}\lambda_{hill}$& $0.302^{+0.061}_{-0.047}$&$0.278$& $0.302^{+0.063}_{-0.049}$ &$0.288$   & $0.35^{+0.12}_{-0.09}$  &$0.342$ \\

$c_{hill}$& $<0.00482$&$0.0036$& $0.0025^{+0.0016}_{-0.0017}$ &$0.0036$   & $0.0046^{+0.0032}_{-0.0031}$  &$0.0055$ \\

$r$ &  $<0.0928$ &$0.016$&  $<0.0941$ &$0.0013$ & $<0.197$&$0.057$\\

$N_{\rm eff}$& $3.18^{+0.58}_{-0.57}$&$3.00$& $3.18^{+0.47}_{-0.44}$ &$3.19$   & $2.74^{+0.64}_{-0.60}$  &$2.52$ \\

$H_0$ &      $68.5 ^{+5.0}_{-4.9}$&$66.9$&      $ 68.5^{+3.0}_{-2.9}$ &$68.3$& $ 64.1^{+5.5}_{-5.4}$ &$62.5$    \\

$\sigma_8$   & $ 0.836 ^{+0.042}_{-0.042}$ &$0.826$  & $ 0.836^{+0.040}_{-0.038}$ &$0.839$  & $ 0.809^{+0.026}_{-0.026}$ &$0.808$ \\

$\chi^2$   && $ 11268.0$  & & $ 11271.7$   && $ 770.4$  \\

\hline
     {\bf Planck TTTEEE}    & & {\bf Best Fit} &&  {\bf Best Fit} &&  {\bf Best Fit} \\                     
 &  {\bf + lowP}   &  {\bf + lowP}  &        {\bf + lowP + BAO } &  {\bf + lowP + BAO} &  {\bf + tau055} &  {\bf + tau055} \\

$\Omega_{\textrm{b}}h^2$& $0.02219^{+0.00049}_{-0.00046} $&$0.02216$& $0.02230^{+0.00038}_{-0.00037} $   &$0.02238$ & $0.02195^{+0.00045}_{-0.00046} $  &$0.02192$ \\

$\Omega_{\textrm{c}}h^2$& $0.1192 ^{+0.0062}_{-0.0059}$&$0.1194$& $0.1196^{+0.0061}_{-0.0061}$  &$0.1189$  & $0.1177 ^{+0.0063}_{-0.0060}$  &$0.1154$ \\

$\tau$& $0.077 \pm 0.035$&$0.067$& $0.082^{+0.030}_{-0.031}$   &$0.087$ & $0.059\pm0.017$ &$0.054$  \\

$10^{12}V_0/M^4$& $<10.5$&$0.25$& $<12.0$  &$1.5$  & $<24.4$  &$5.7$ \\

$log(b[GeV])$& $>6.85$&$8.4$& $>6.77$   &$9.0$ & $>7.27$  &$18.1$ \\

$10^{11}\lambda_{hill}$& $0.303^{+0.055}_{-0.044}$&$0.269$& $0.306^{+0.061}_{-0.047}$ &$0.288$   & $0.34^{+0.11}_{-0.08}$ &$0.316$  \\

$c_{hill}$& $0.0035^{+0.0018}_{-0.0018}$&$0.0035$& $0.0030^{+0.0015}_{-0.0014}$  &$0.0033$  & $0.0042^{+0.0020}_{-0.0019}$  &$0.0046$ \\

$r$ &  $<0.0841$ &$0.0022$&  $<0.0957$  &$0.013$& $<0.182$&$0.047$\\

$N_{\rm eff}$& $2.99^{+0.41}_{-0.39}$&$3.01$& $3.07^{+0.37}_{-0.36}$   &$3.02$ & $2.81^{+0.42}_{-0.38}$&$2.68$   \\

$H_0$ &      $66.9^{+3.2}_{-3.0}$&$66.9$&      $ 67.7\pm 2.4$ &$67.4$& $ 65.1^{+3.2}_{-3.0}$  &$64.4$   \\

$\sigma_8$   & $ 0.827 ^{+0.036}_{-0.034}$  &$0.822$ & $ 0.832\,\pm0.032$   &$0.834$& $ 0.810\,\pm 0.024$ &$0.799$ \\

$\chi^2$   && $ 12946.0$   && $ 12948.3$   && $ 2447.9$  \\

\hline

\end{tabular}}

\end{table}

Regarding the inflationary parameters in this extended $\Lambda$CDM+r+$N_{\rm eff}$ scenario, we still predict a tensor-to-scalar ratio consistent with zero, i.e., $r<0.0928$ for Planck TT+lowP. Moreover, $b$ still has a lower limit, i.e., $b > 2.6 \times10^6 GeV$ and $V_0<11.7\times 10^{-12} M_P^4$, for Planck TT+lowP datasets.

Finally, in Table~\ref{tablewp4} we show the constraints for the $w$CDM+r scenario, using the Hilltop inflationary model with $p=4$, to test the modifications derived from quantum entanglement from this theory of the origin of the universe.
In our modified Hilltop inflationary model, also by varying the equation of state of the dark energy, we find robust constraints for most of the cosmological parameters, which have no significant differences with respect to the standard $w$CDM model, as shown in the last two columns of Table~\ref{tablelcdm}, and with respect to the original Hilltop model with $p=4$, how can be seen by looking at Table~\ref{tableunmod_p4}. However, also considering these extensions of the model, for our modified Hilltop inflation the $\chi^2$ gets worse. Also, in this case, the mainly reason for extending the baseline scenario is trying to solve the Hubble constant tension, adding a free dark energy equation of state. In fact, the geometrical degeneracy existing between $w$ and $H_0$ that produces a very large shift of the Hubble constant, unconstrained in this scenario, is very well known.
Also, in this modified Hilltop inflation with $p=4$,  as it has been shown by several authors~\cite{DiValentino:2015ola,DiValentino:2016hlg, Qing-Guo:2016ykt,DiValentino:2017iww,DiValentino:2017zyq,Mortsell:2018mfj,Yang:2018euj}, the tension is solved with an equation of state $w<-1$.
In fact, we have with Planck TTTEEE+lowP $w=-1.56\,^{+0.59}_{-0.47}$ and $H_0>66$ Km/s/Mpc, in complete agreement with~\cite{R16}.
However, when we add the BAO dataset, we break their degeneracy and the dark energy equation of state recover the expected value $w=-1$, so a slightly tension at about $2\sigma$ reappears in the Hubble constant estimation.

\begin{table}[]
\caption{$95 \% $ c.l. constraints on cosmological parameters in our baseline $w$CDM+r scenario from different combinations of datasets with a modified Hilltop inflation  with $p=4$.}
\label{tablewp4}
\centering
\scalebox{0.8}{\begin{tabular}{lccccccc}
\hline
      {\bf Planck TT}    & &{\bf Best Fit}&&{\bf Best Fit}&&{\bf Best Fit} \\                     
 & {\bf + lowP}   &{\bf + lowP} &        {\bf + lowP + BAO} &{\bf + lowP + BAO}&{\bf + tau055}&{\bf + tau055} \\

$\Omega_{\textrm{b}}h^2$& $0.02228^{+0.00048}_{-0.00044} $&$0.02236$& $0.02225^{+0.00043}_{-0.00041}$  &$0.02249$  & $0.02215\pm 0.00042 $ &$0.02222$  \\

$\Omega_{\textrm{c}}h^2$& $0.1193^{+0.0043}_{-0.0045}$&$0.1193$& $0.1192\pm 0.0038$  &$0.1176$  & $0.1210^{+0.0041}_{-0.0042} $  &$0.1211$ \\

$\tau$& $0.076^{+0.039}_{-0.037}$&$0.086$& $0.078^{+0.037}_{-0.037}$   &$0.078$ & $0.058\,^{+0.017}_{-0.016}$  &$0.054$ \\

$10^{12}V_0/M^4$& $<12.5$&$1.6$& $<12.0$  &$3.8$  & $<28.3$ &$1.2$  \\

$log(b[GeV])$& $>6.91$&$7.4$& $>6.70$    &$17.0$& $>7.12$ &$14.4$  \\

$10^{11}\lambda_{hill}$& $0.306^{+0.063}_{-0.050}$&$0.284$& $0.303^{+0.062}_{-0.049}$ &$0.294$   & $0.35^{+0.12}_{-0.10}$   &$0.273$\\

$c_{hill}$& $0.0030^{+0.0012}_{-0.0012}$&$0.0030$& $0.0030\pm 0.0011$  &$0.0024$  & $0.0032^{+0.0012}_{-0.0012}$ &$0.0038$  \\

$r$ &  $<0.100$ &$0.013$&  $<0.0972$&$0.033$  & $<0.210$&$0.010$\\

$w$& $-1.53^{+0.60}_{-0.49}$&$-1.78$& $-1.02^{+0.15}_{-0.15}$  &$-0.97$  & $-1.48^{+0.69}_{-0.60}$  &$-1.68$ \\

$H_0$ &      $>65$&$94.1$&      $ 68.1^{+3.4}_{-3.3}$ &$67.3$& $ >62$  &$88.3$   \\

$\sigma_8$   & $ 0.98 ^{+0.14}_{-0.17}$  &$1.06$ & $ 0.834^{+0.053}_{-0.050}$  &$0.812$ & $ 0.95^{+0.16}_{-0.19}$ &$1.00$ \\

$\chi^2$   && $11262.7$   && $ 11272.6$   && $ 769.9$  \\

\hline

     {\bf Planck TTTEEE}    & & {\bf Best Fit} &&  {\bf Best Fit} &&  {\bf Best Fit} \\                     
 &  {\bf + lowP}   &  {\bf + lowP}  &        {\bf + lowP + BAO } &  {\bf + lowP + BAO} &  {\bf + tau055} &  {\bf + tau055} \\

$\Omega_{\textrm{b}}h^2$& $0.02227^{+0.00031}_{-0.00030} $&$0.02231$& $0.02225^{+0.00030}_{-0.00029}  $   &$0.02230$ & $0.02218^{+0.00029}_{-0.00028} $ &$0.02231$  \\

$\Omega_{\textrm{c}}h^2$& $0.1196 ^{+0.0029}_{-0.0028}$&$0.1195$& $0.1197^{+0.0029}_{-0.0027}$  &$0.1195$  & $0.1206 ^{+0.0028}_{-0.0028}$ &$0.1201$  \\

$\tau$& $0.074 ^{+0.032}_{-0.034}$&$0.082$& $0.078\pm0.033$  &$0.090$  & $0.060^{+0.017}_{-0.017}$ &$0.055$  \\

$10^{12}V_0/M^4$& $<13.5$&$0.12$& $<11.9$   &$0.62$ & $<23.4$  &$5.5$ \\

$log(b[GeV])$& $>6.96$&$13.5$& $>6.93$  &$14.6$  & $>7.40$  &$14.5$ \\

$10^{11}\lambda_{hill}$& $0.308^{+0.066}_{-0.051}$&$0.275$& $0.306^{+0.061}_{-0.048}$ &$0.281$   & $0.343^{+0.097}_{-0.087}$  &$0.307$ \\

$c_{hill}$& $0.0031^{+0.0010}_{-0.0009}$&$0.0034$& $0.00320^{+0.00096}_{-0.00096}$  &$0.0032$  & $0.0032\pm 0.0010$  &$0.0034$ \\

$r$ &  $<0.107$ &$0.0010$&  $<0.0947$  &$0.0052$& $<0.178$&$0.046$\\

$w$& $-1.56^{+0.59}_{-0.47}$&$-1.36$& $-1.03^{+0.11}_{-0.13}$   &$-1.04$ & $-1.54^{+0.65}_{-0.52}$&$-1.48$   \\

$H_0$ &      $>66$&$78.5$&      $ 68.3^{+3.2}_{-2.9}$ &$68.8$& $ >64$   &$82.4$  \\

$\sigma_8$   & $ 0.99 ^{+0.13}_{-0.17}$  &$0.934$ & $ 0.839^{+0.044}_{-0.040}$ &$0.852$  & $ 0.97^{+0.14}_{-0.18}$&$0.946$  \\

$\chi^2$   && $12941.8$   && $ 12950.7$   && $ 2447.8$  \\

\hline

\end{tabular}}

\end{table}

Regarding the inflationary parameters, again we find just un upper limit for the tensor-to-scalar ratio, i.e., $r<0.107$, and just a lower limit for the 'SUSY-breaking' scale associated with the landscape effects $b$, i.e., $b>9.1\times10^6 GeV$ and $V_0<13.5\times 10^{-12} M_P^4$ for Planck TT+lowP. 

\begin{figure}[]
\centering
\includegraphics[width=7.5cm]{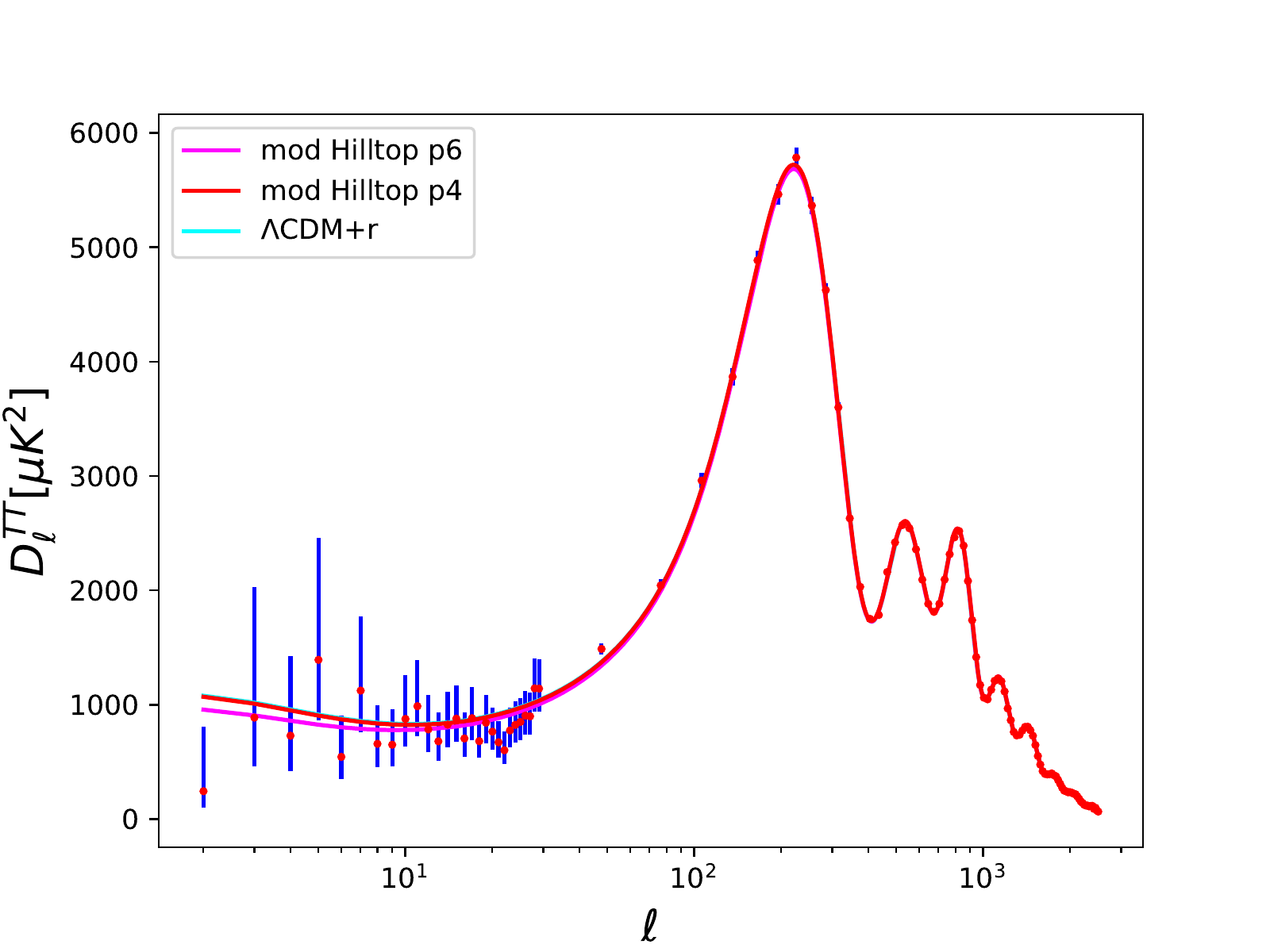}
\includegraphics[width=7.5cm]{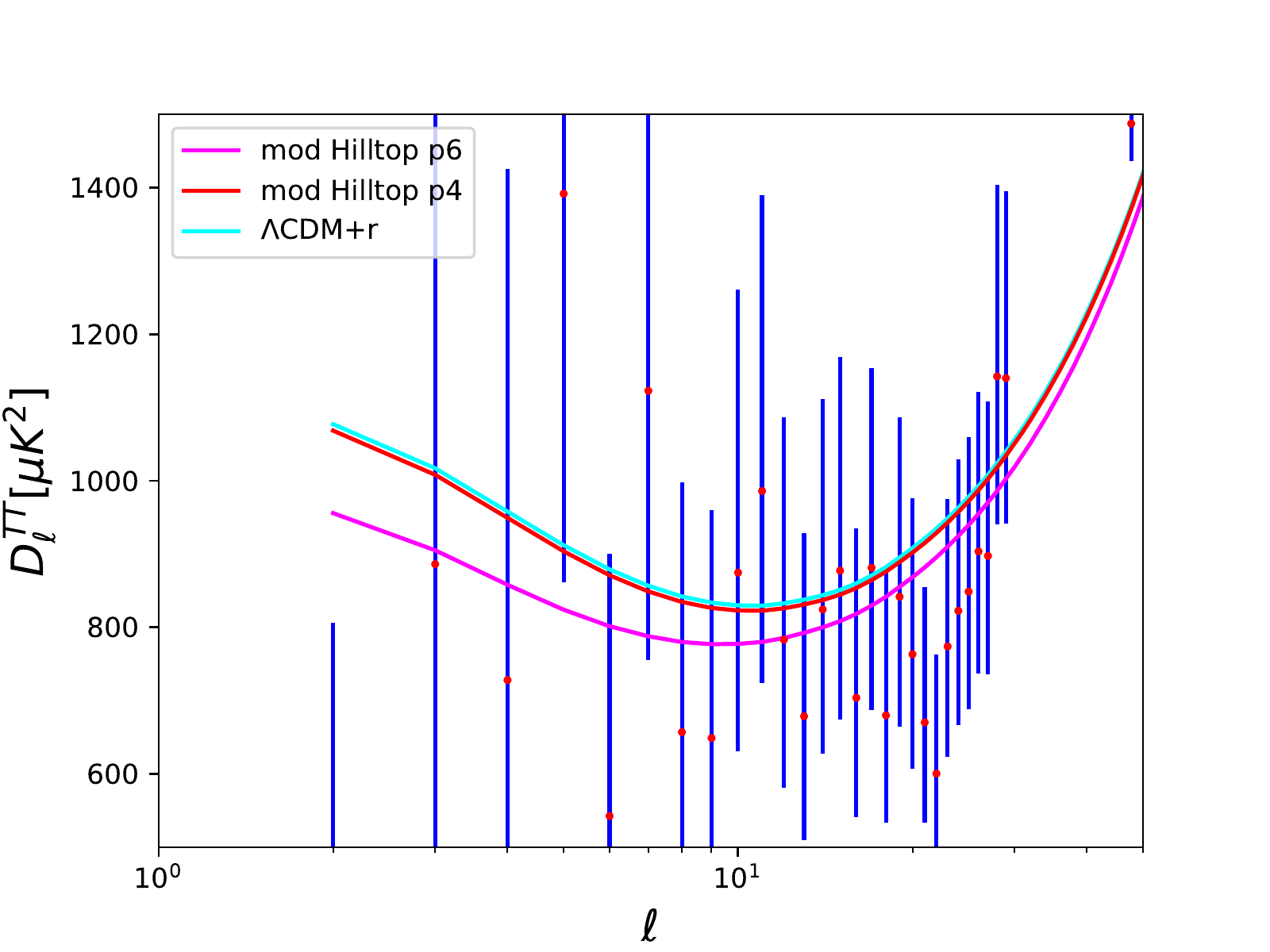}
\caption{Comparison of the temperature CMB angular power spectrum computed for the best fit of our modified Hilltop model with $p=6$ (magenta), the best fit of our modified Hilltop model with $p=4$ (red), and the best fit obtained with a minimal standard cosmological model $\Lambda$CDM+r (cyan), with Planck 2015 TT+lowP data (points with error bars). The main differences between the two models are at lower-$\ell$ and on the amplitude of the peaks that the Hilltop model with $p=6$, modified for the entanglement, prefers slightly lower.}
\label{clbf}
\end{figure}

\begin{figure}[]
\centering
\includegraphics[width=7.5cm]{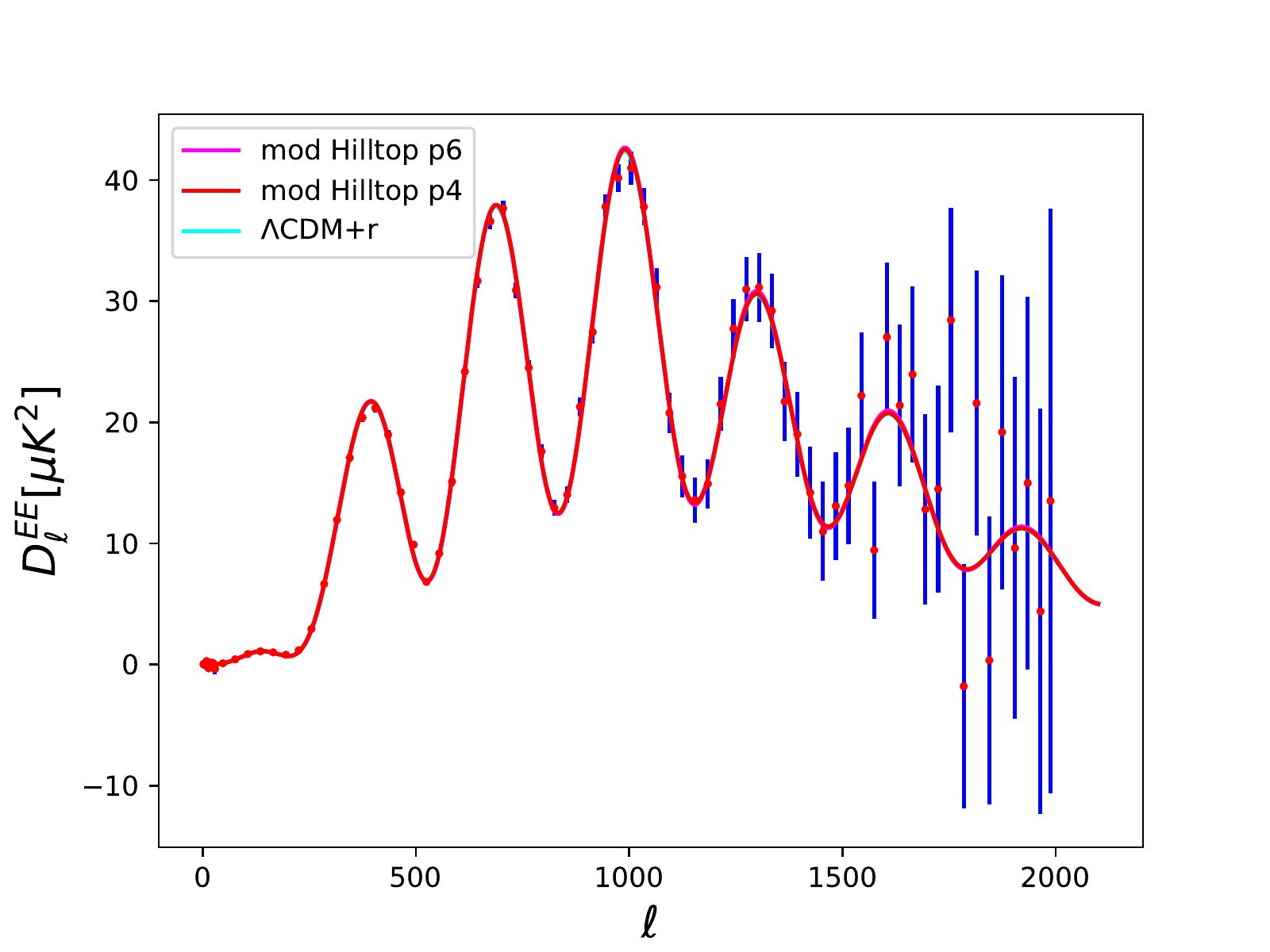}
\includegraphics[width=7.5cm]{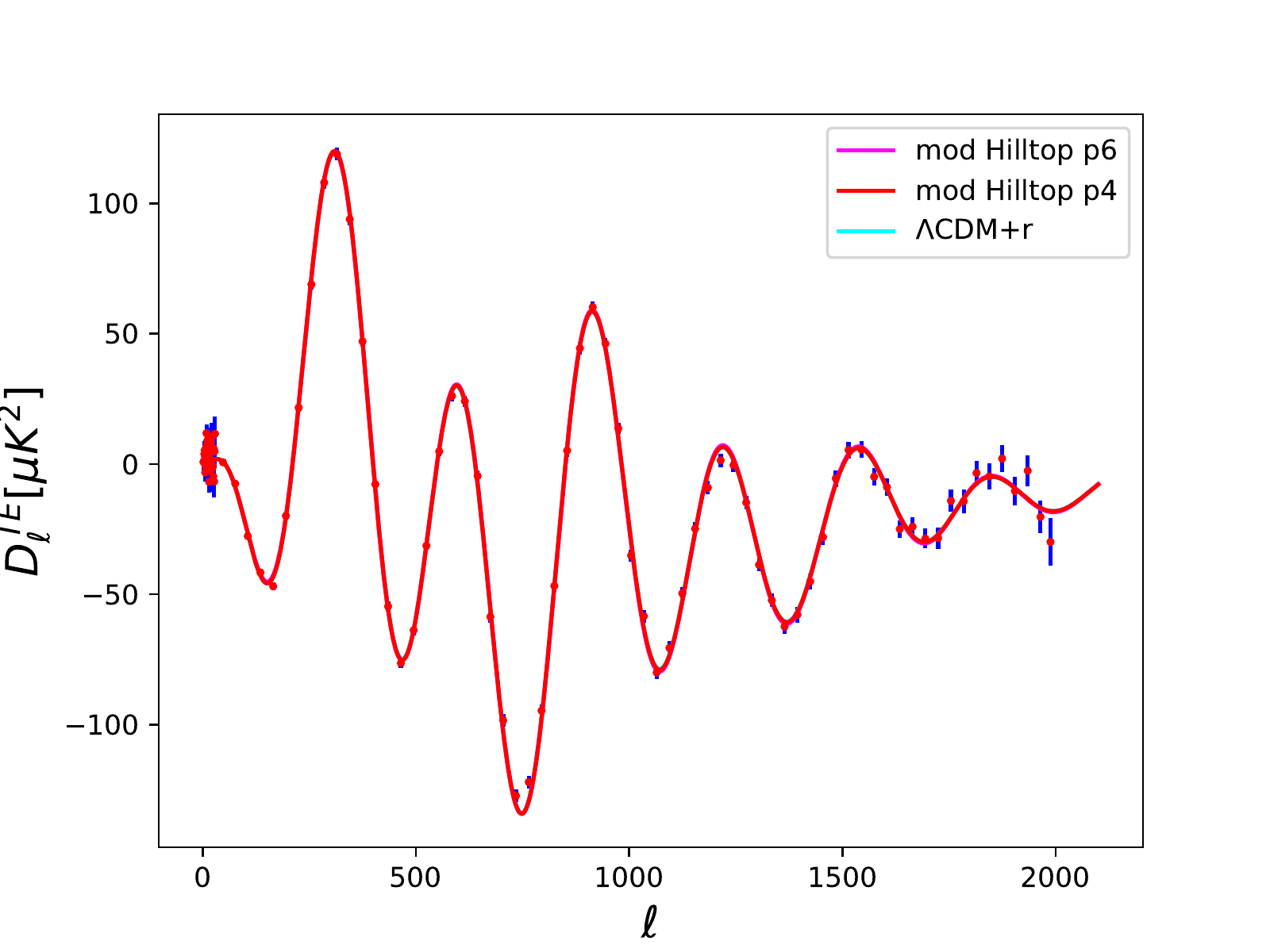}
\caption{Comparison of the polarization CMB angular power spectra computed for the best fit of our modified Hilltop model with $p=6$ (magenta), the best fit of our modified Hilltop model with $p=4$ (red), and the best fit obtained with a minimal standard cosmological model $\Lambda$CDM+r (cyan), with Planck 2015 TT+lowP data (points with error bars).}
\label{clbf_pol}
\end{figure}

If we look at the Figures~\ref{clbf} and \ref{clbf_pol}, we can see the temperature and polarization power spectra obtained with the best fit of our modified Hilltop model with $p=4$ and the best fit of a minimal standard cosmological model $\Lambda$CDM+r, compared with Planck 2015 TT+lowP data: they are about indistinguishable, fitting the data in the same manner.

\section{Results  $p=6$}\label{resultsp6}

The result of all the explorations assuming a modified Hilltop scenario with $p=6$, are given in Tables~\ref{tabler}--\ref{tablew}, where we report the constraints at $95 \% $ c.l. on the cosmological parameters, respectively for the $\Lambda$CDM+r, $\Lambda$CDM+r+$N_{\rm eff}$, $w$CDM+r models. In Table~\ref{tableunmod_p6} we can see instead the bounds for the same cases for the unmodified Hilltop scenario with $p=6$.

\begin{figure}[]
\centering
\includegraphics[scale=0.6]{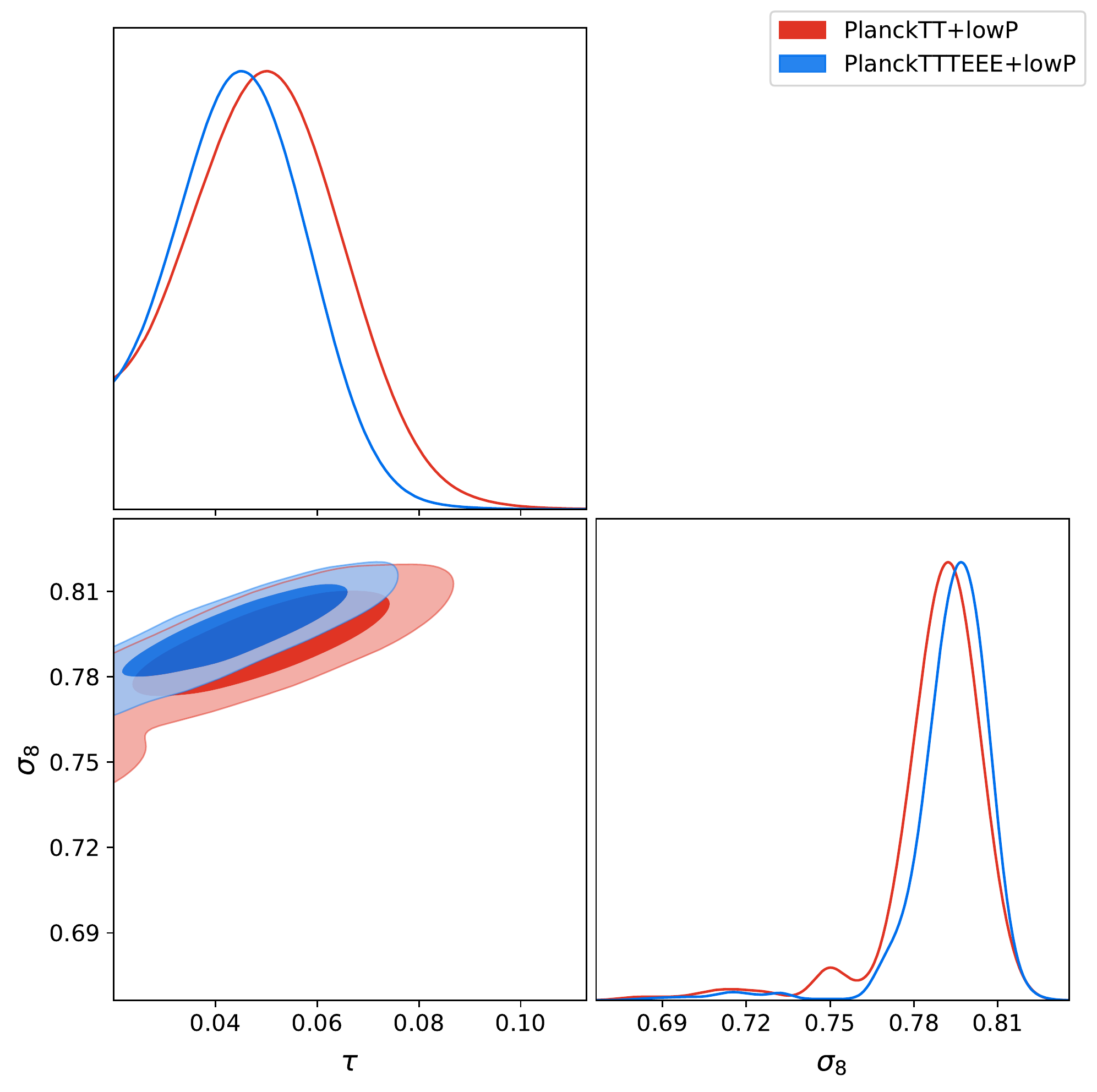}
\caption{Constraints at $68 \%$ and  $95 \%$ confidence levels on the $\sigma_8$ vs. $\tau$ plane, in our modified $\Lambda$CDM+r Hilltop inflation with $p=6$. Looking at the Table~\ref{tabler}, we can see that the best fits for these parameters are $\sigma_8=0.793$ and $\tau=0.043$ for PlanckTT+lowP, while they are $\sigma_8=0.798$ and $\tau=0.044$ for PlanckTTTEEE+lowP.}
\label{figs8tau}
\end{figure}

\begin{figure}[]
\centering
\includegraphics[width=7cm]{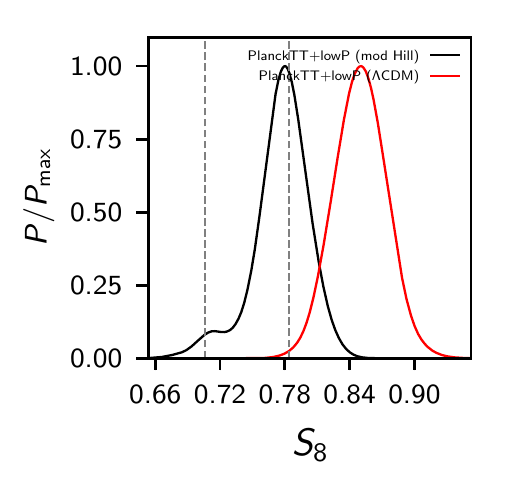}
\caption{1D posteriors for our modified $\Lambda$CDM+r Hilltop inflation model with $p=6$ (black solid line) and the standard $\Lambda$CDM model (red solid line). The region between the grey dashed lines is the $1\sigma$ constraint obtained by KiDS-450.}
\label{S8}
\end{figure}

If we compare Table~\ref{tabler}, where there are the constraints for this modified Hilltop inflation with $p=6$, and the constraints obtained in Table~\ref{tablerp4}, where the are those for $p=4$, we see a very large shift for most of the cosmological parameters. In particular, looking at PlanckTT+lowP we see an important shift of $\Omega_ch^2$, $\tau$ and $\sigma_8$ at about $2\sigma$ towards lower values (see Figure~\ref{figs8tau}), and of $H_0$ of about $1\sigma$ towards a higher one. However, these shifts are not due to our modifications but are characteristic of the Hilltop model with $p=6$ itself, how can be appreciated by looking at Table~\ref{tableunmod_p6}. In any case, all these shifts are interesting because seem to go in the right direction for solving the several tensions we see in the cosmological data, between Planck and the other experiments. For example, the well-known degeneracy between the Planck satellite~\cite{planckparams2013,planckparams2015,planckparams2018} and the local measurements of the Hubble constant of Riess et al.~\cite{R11,R16,R18}, in this case decreased at $2.2\sigma$. Moreover, the tension between Planck and the weak lensing experiments such as the Canada France Hawaii Lensing Survey (CFHTLenS)~\cite{Heymans:2012gg, Erben:2012zw}, the Kilo Degree Survey of$~$450 deg$^2$ of imaging data (KiDS-450)~\cite{Hildebrandt:2016iqg}, and the Dark Energy Survey (DES)~\cite{Abbott:2017wau}, about the $S_8 \equiv\sigma_8 \sqrt{\Omega_m/0.3}$ value. Thanks to the fact that both the matter density and the clustering parameter are going down in the modified Hilltop model, we find $S_8=0.776\pm0.027$ at $68\%$ c.l., reducing for example the tension with the value $S_8=0.745\pm0.039$ at $68\%$ c.l. measured by KiDS-450~\cite{Hildebrandt:2016iqg} within $1\sigma$, as we can see in Figure~\ref{S8}. Finally, the reionization optical depth obtained is now shifted towards lower values, perfectly in agreement with the new $\tau=0.055\pm0.009$ at $68\%$ c.l. obtained from Planck HFI measurements~\cite{newtau} and released in the new Planck 2018 parameters paper~\cite{planckparams2018}. However, for our modified Hilltop inflation the $\chi^2$ gets worse of about $20$.

\begin{table}[]
\caption{$95 \% $ c.l. constraints on cosmological parameters in our baseline $\Lambda$CDM+r scenario from different combinations of datasets with a modified Hilltop inflation with $p=6$.}
\label{tabler}
\centering
\scalebox{0.9}{\begin{tabular}{lccccccc}
\hline
     {\bf Planck TT}    & &{\bf Best Fit}&&{\bf Best Fit}&&{\bf Best Fit} \\                     
 & {\bf + lowP}   &{\bf + lowP} &        {\bf + lowP + BAO} &{\bf + lowP + BAO}&{\bf + tau055}&{\bf + tau055} \\

$\Omega_{\textrm{b}}h^2$& $0.02232^{+0.00046}_{-0.00045} $&$0.02229$& $0.02222\,\pm0.00039$    &$0.02222$& $0.02222^{+0.00044}_{-0.00043} $  &$0.02209$ \\

$\Omega_{\textrm{c}}h^2$& $0.1158^{+0.0040}_{-0.0038}$&$0.1172$& $0.1175\,^{+0.0024}_{-0.0025}$  &$0.1179$  & $0.1181^{+0.0038}_{-0.0036} $  &$0.1191$ \\

$\tau$& $0.048^{+0.030}_{-0.031}$&$0.043$& $0.042\,^{+0.026}_{-0.029}$ &$0.051$   & $0.055\,\pm 0.017$  &$0.057$ \\

$10^{12}V_0/M^4$& $48 \,\pm 10$&$46$& $49\,\pm10$  &$48$  & $60\,\pm10$ &$56$  \\

$log(b[GeV])$& $>8.10$&$11.7$& $>8.01$ &$16.8$   & $>8.30$ &$8.0$  \\

$10^{11}\lambda_{hill}$& $>0.951$&$0.999$& $>0.954$   &$0.999$ & $>0.936$ &$0.998$  \\

$c_{hill}$& $0.00178\,^{+0.00095}_{-0.00098}$&$0.0021$& $0.00211^{+0.00070}_{-0.00074}$  &$0.0020$  & $0.00234^{+0.00080}_{-0.00083}$  &$0.0025$ \\

$r$ &  $0.398\,\pm 0.065$ &$0.38$&  $0.401\,^{+0.065}_{-0.064}$  &$0.39$& $0.469\,^{+0.087}_{-0.077}$&$0.443$\\

$H_0$ &      $68.9 \pm 1.8$&$68.3$&      $ 68.2\pm1.1$ &$68.1$& $ 67.9\,\pm1.7$   &$67.4$  \\

$\sigma_8$   & $ 0.791\,^{+0.022}_{-0.023}$   &$0.793$& $ 0.793\,\pm0.021$ &$0.800$  & $ 0.806^{+0.016}_{-0.017}$ &$0.810$ \\

$\chi^2$   && $11296.3$   && $ 11302.4$   && $ 781.8$  \\
\hline

    {\bf Planck TTTEEE}    & & {\bf Best Fit} &&  {\bf Best Fit} &&  {\bf Best Fit} \\                     
 &  {\bf + lowP}   &  {\bf + lowP}  &        {\bf + lowP + BAO } &  {\bf + lowP + BAO} &  {\bf + tau055} &  {\bf + tau055} \\

$\Omega_{\textrm{b}}h^2$& $0.02226^{+0.00031}_{-0.00032} $&$0.02213$& $0.02224^{+0.00027}_{-0.00028} $  &$0.02228$  & $0.02222\,\pm0.00029 $ &$0.02216$  \\

$\Omega_{\textrm{c}}h^2$& $0.1179 \, ^{+0.0028}_{-0.0027}$&$0.1183$& $0.1182^{+0.0020}_{-0.0021}$    &$0.1182$& $0.1189\,\pm0.0026 $   &$0.1196$\\

$\tau$& $0.044\,^{+0.026}_{-0.027}$&$0.044$& $0.043\,\pm0.026$  &$0.048$  & $0.054\,\pm0.016$ &$0.054$  \\

$10^{12}V_0/M^4$& $49\,\pm10$&$48$& $50\,\pm10$  &$50$  & $60\,\pm10$  &$59$ \\

$log(b[GeV])$& $>8.16$&$18.8$& $>8.24$   &$16.5$ & $>7.93$ &$10.6$  \\

$10^{11}\lambda_{hill}$& $>0.964$&$0.996$& $>0.962$  &$0.995$  & $>0.957$ &$0.999$  \\

$c_{hill}$& $0.00219^{+0.00074}_{-0.00072}$&$0.0023$& $0.00224^{+0.00065}_{-0.00066}$  &$0.0021$  & $0.00249^{+0.00065}_{-0.00066}$ &$0.0027$  \\

$r$ &  $0.405\,^{+0.057}_{-0.056}$ &$0.397$&  $0.406\,^{+0.059}_{-0.058}$  &$0.40$& $0.463\,^{+0.068}_{-0.064}$&$0.45$\\

$H_0$ &      $68.0^{+1.2}_{-1.3}$&$67.8$&      $ 67.88^{+0.95}_{-0.91}$ &$67.9$& $ 67.6 \,\pm1.2$    &$67.3$ \\

$\sigma_8$   & $ 0.796\,^{+0.019}_{-0.020}$   &$0.798$& $ 0.796\,^{+0.020}_{-0.019}$  &$0.802$ & $ 0.808\,\pm 0.015$ &$0.811$ \\

$\chi^2$   && $12981.0$   && $ 12986.4$   && $ 2466.4$  \\

\hline
\end{tabular}}

\end{table}

\begin{figure}[]
\centering
\includegraphics[scale=0.6]{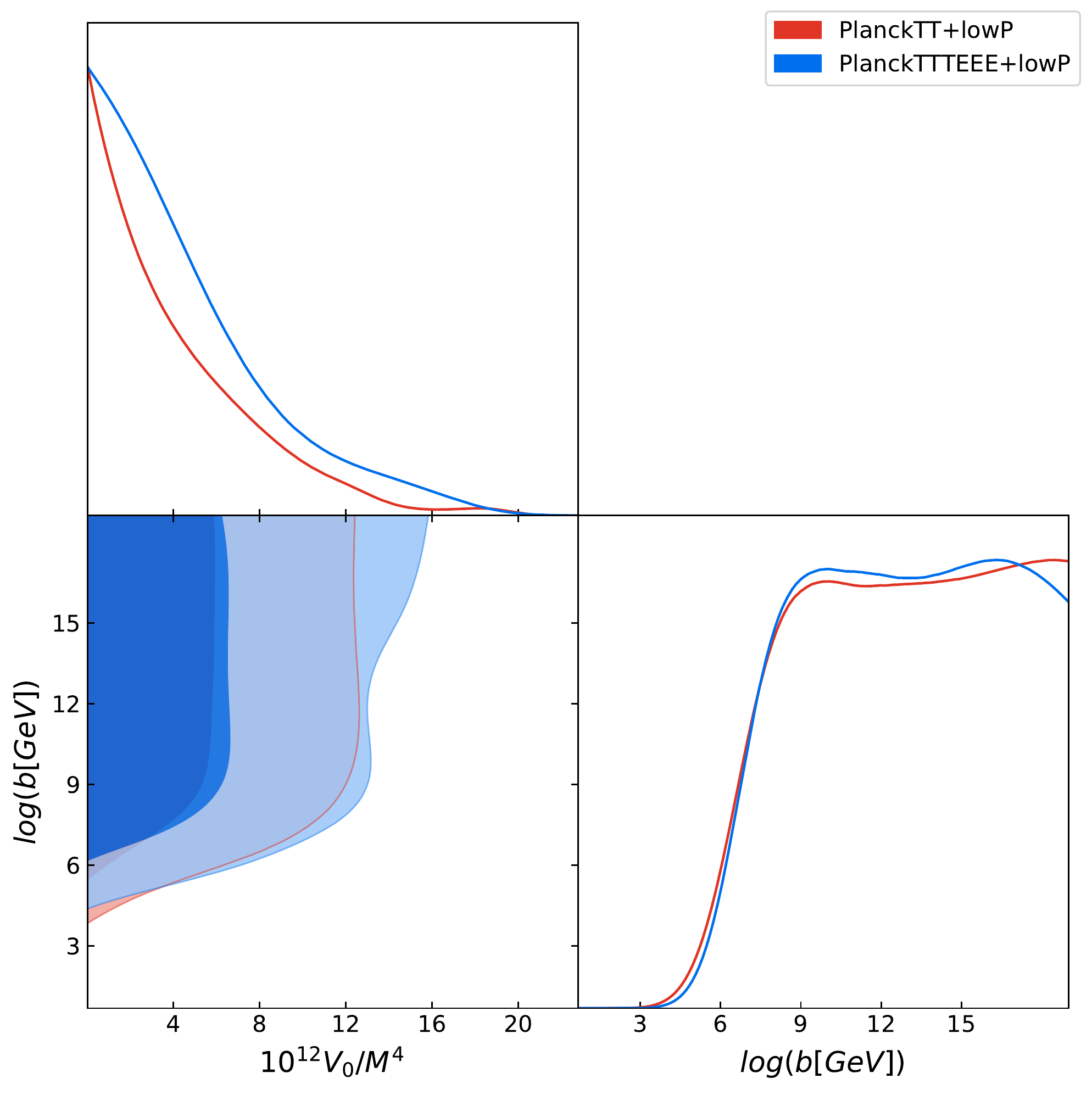}
\caption{Constraints at $68 \%$ and  $95 \%$ confidence levels on the $12^{10}V_0/M^4$ vs. $Log(b[GeV])$ plane, in our modified $\Lambda$CDM+r Hilltop inflation with $p=6$. Looking at the Table~\ref{tabler}, we can see that the best fits for these parameters are $12^{10}V_0/M^4 = 46$ and $log(b[GeV])=11.7$ for PlanckTT+lowP, while they are $12^{10}V_0/M^4 = 48$ and $log(b[GeV])=18.8$ for PlanckTTTEEE+lowP.}
\label{figv0b}
\end{figure}

Regarding the inflationary parameters that describe the theory analyzed here, we have now a prediction for the tensor-to-scalar ratio $r$, which in our analysis is a derived parameter, different from zero at many standard deviations. We find for this model and Planck TT + lowP that $r=0.398\,\pm0.065$, and probably is this value not supported by the data to worsen the $\chi^2$ value. If we look at Figure~\ref{figv0b}, which shows the constraints at $68 \%$ and  $95 \%$ confidence levels on the $10^{12}V_0/M^4$ vs. $log(b)$ plane, we can see that there exists a lower limit for $b$ at $b>1.3\times10^8 GeV$ stronger than the $p=4$ case and $V_0=(48\pm10)\times 10^{-12} M_P^4$, shifted towards higher values with respect to the $p=4$ case, for Planck TT+lowP. Finally, we pass from the detection of a value of $\lambda_{hill}=0.303^{+0.059}_{-0.045}$ in Table~\ref{tablerp4} to just a lower limit $\lambda_{hill}>0.951$ in Table~\ref{tabler}, and from $c_{hill}=0.0031\pm0.0012$ to $c_{hill}=0.00178^{+0.00095}_{-0.00098}$ for PlanckTT+lowP.

The same conclusions arise by adding the polarization data of Planck at high-$\ell$, the BAO data or by using the ``tau055'' prior, confirming the robustness of our results.

\begin{figure}[]
\centering
\includegraphics[scale=0.7]{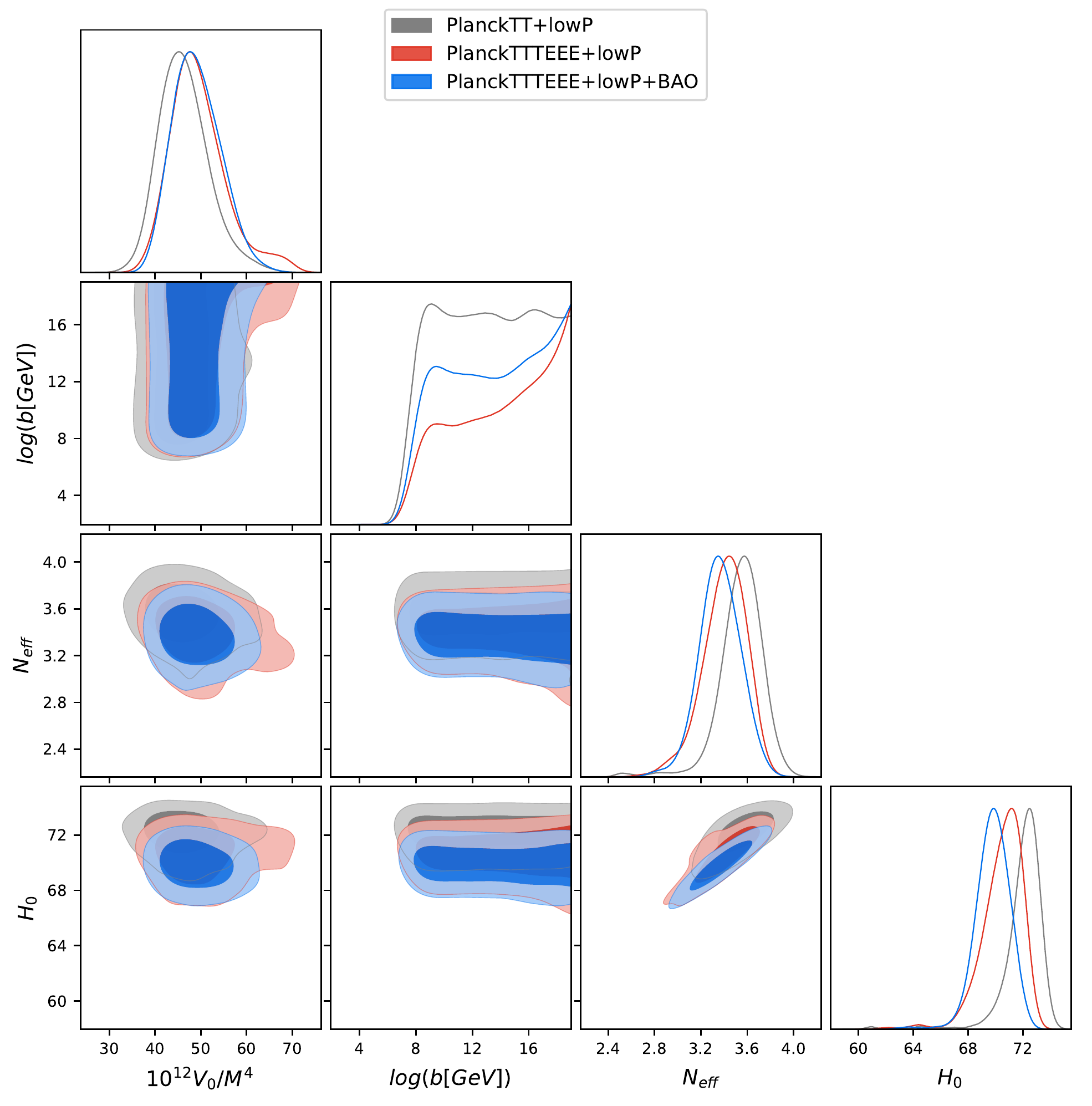}
\caption{Constraints at $68 \%$ and  $95 \%$ confidence levels in our modified $\Lambda$CDM+r+$N_{\rm eff}$ Hilltop scenario with $p=6$. Looking at the Table~\ref{tablennu}, we can see that the best fits for these parameters are $12^{10}V_0/M^4 = 40$, $log(b[GeV])=11$, $N_{\rm eff}=3.66$ and $H_0=73.1$ for PlanckTT+lowP, while they are $12^{10}V_0/M^4 = 48$, $log(b[GeV])=18.9$, $N_{\rm eff}=3.61$ and $H_0=71.8$ for PlanckTTTEEE+lowP.}
\label{fignnu}
\end{figure}

In addition, if we compare Table~\ref{tablennu}, where there are the constraints for this modified Hilltop scenario with $p=6$, by introducing a dark radiation component free to vary $N_{\rm eff}$, and the constraints obtained in the same scenario for $p=4$ in Table~\ref{tablennup4}, we see a similar shift of the cosmological parameters than in the $\Lambda$CDM+r case. However, also in this case these shifts are not related to our modifications but to the Hilltop model with $p=6$ itself, how can be seen by looking at Table~\ref{tableunmod_p6}. 
In particular, looking at PlanckTT+lowP we see a shift of $\Omega_ch^2$, $\tau$ and $\sigma_8$ at about $1\sigma$ towards lower values, and of $H_0$ of more than $1\sigma$ towards a higher one, so always in the direction of solving the tensions between the different cosmological probes. In this case, the Hubble constant tension is solved within $1\sigma$, thanks to the evidence for a dark radiation $N_{eff}>3.045$ at about $3\sigma$. On the contrary of what usually happens, this evidence is slightly reduced, but not disappears, even when we consider PlanckTTTEEE+lowP, see Figure~\ref{fignnu}. Therefore, also when the polarization of Planck is added, we can solve the disagreement on the Hubble constant between the CMB and the direct measurements by considering a dark radiation component, as we found also in~\cite{eleonoraE}. Also, in this case the $\chi^2$ value of our modified Hilltop inflation gets worse of about $20$ with respect to the standard inflationary model, but it performs about $15$ better than the original Hilltop inflation for the Planck TT+lowP case.

\begin{table}[]
\caption{$95 \% $ c.l. constraints on cosmological parameters in our baseline $\Lambda$CDM+r+$N_{\rm eff}$ scenario from different combinations of datasets with a modified Hilltop inflation  with $p=6$.}
\label{tablennu}
\centering
\scalebox{0.9}{\begin{tabular}{lccccccc}
\hline
     {\bf Planck TT}    & &{\bf Best Fit}&&{\bf Best Fit}&&{\bf Best Fit} \\                     
 & {\bf + lowP}   &{\bf + lowP} &        {\bf + lowP + BAO} &{\bf + lowP + BAO}&{\bf + tau055}&{\bf + tau055} \\

$\Omega_{\textrm{b}}h^2$& $0.02236^{+0.00041}_{-0.00044} $&$0.02278$& $0.02248^{+0.00043}_{-0.00044}$   &$0.02262$ & $0.02248^{+0.00049}_{-0.00052} $  &$0.02234$ \\

$\Omega_{\textrm{c}}h^2$& $0.1236^{+0.0068}_{-0.0067}$&$0.1245$& $0.1256^{+0.0071}_{-0.0073}$  &$0.1283$  & $0.1231^{+0.0074}_{-0.0073} $ &$0.1212$  \\

$\tau$& $0.055^{+0.026}_{-0.030}$&$0.051$& $0.049^{+0.028}_{-0.029}$  &$0.067$  & $0.058\,^{+0.016}_{-0.017}$&$0.056$   \\

$10^{12}V_0/M^4$& $46\,\pm10$&$40$& $47\,\pm10$   &$46$ & $55\,^{+20}_{-10}$  &$51$ \\

$log(b[GeV])$& $>7.83$&$11.0$& $>7.75$   &$9.5$ & $>7.55$ &$16.8$  \\

$10^{11}\lambda_{hill}$& $>0.937$&$0.995$& $>0.940$  &$0.998$  & $>0.916$  &$0.993$ \\

$c_{hill}$& $<0.00127$&$0.0000$& $<0.00185$   &$0.0000$ & $<0.00256$  &$0.0017$ \\

$r$ &  $0.388\,^{+0.068}_{-0.061}$ &$0.35$&  $0.392\,^{+0.071}_{-0.069}$ &$0.38$ & $0.446\,^{+0.097}_{-0.086}$&$0.41$\\

$N_{\rm eff}$& $3.56\,^{+0.30}_{-0.34}$&$3.66$& $3.54^{+0.36}_{-0.41}$  &$3.73$  & $3.42^{+0.44}_{-0.46}$ &$3.28$  \\

$H_0$ &      $72.1 ^{+2.0}_{-2.3}$&$73.1$&      $ 71.0\,^{+2.3}_{-2.4}$&$72.2$ & $ 70.6^{+3.2}_{-3.5}$ &$69.5$    \\

$\sigma_8$   & $ 0.819 ^{+0.027}_{-0.028}$  &$0.817$ & $ 0.821^{+0.032}_{-0.033}$  &$0.846$ & $ 0.821^{+0.026}_{-0.027}$ &$0.815$ \\

$\chi^2$   && $11287.9$   && $ 11298.3$  & & $ 781.3$  \\

\hline

     {\bf Planck TTTEEE}    & & {\bf Best Fit} &&  {\bf Best Fit} &&  {\bf Best Fit} \\                     
 &  {\bf + lowP}   &  {\bf + lowP}  &        {\bf + lowP + BAO } &  {\bf + lowP + BAO} &  {\bf + tau055} &  {\bf + tau055} \\

$\Omega_{\textrm{b}}h^2$& $0.02258\,^{+0.00040}_{-0.00042} $&$0.02275$& $0.02248\,^{+0.00036}_{-0.00037} $ &$0.02246$   & $0.02236\,\pm0.00042 $  &$0.02216$ \\

$\Omega_{\textrm{c}}h^2$& $0.1233 ^{+0.0057}_{-0.0058}$&$0.1257$& $0.1235^{+0.0059}_{-0.0057}$   &$0.1210$ & $0.1213 ^{+0.0061}_{-0.0058}$ &$0.1180$  \\

$\tau$& $0.054 \,^{+0.028}_{-0.030}$&$0.070$& $0.050\,^{+0.026}_{-0.029}$  &$0.062$  & $0.056\,^{+0.017}_{-0.016}$ &$0.053$  \\

$10^{12}V_0/M^4$& $48\,\pm10$&$48$& $48\,\pm10$  &$52$  & $58\,\pm10$ &$56$  \\

$log(b[GeV])$& $>7.99$&$18.9$& $>8.02$  &$8.4$  & $>8.00$ &$8.5$  \\

$10^{11}\lambda_{hill}$& $>0.955$&$0.997$& $>0.956$  &$0.994$  & $>0.950$ &$0.994$  \\

$c_{hill}$& $<0.00212$&$0.0002$& $0.0013^{+0.0011}_{-0.0012}$  &$0.0015$  & $0.0020^{+0.0012}_{-0.0013}$ &$0.0026$  \\

$r$ &  $0.398\,^{+0.063}_{-0.060}$ &$0.39$&  $0.400\,^{+0.058}_{-0.055}$  &$0.42$& $0.453\,^{+0.076}_{-0.071}$&$0.44$\\

$N_{\rm eff}$& $3.42\,^{+0.33}_{-0.35}$&$3.61$& $3.38\pm0.34$  &$3.26$  & $3.21^{+0.38}_{-0.35}$&$3.00$   \\

$H_0$ &      $70.6\,^{+2.4}_{-2.6}$&$71.8$&      $ 69.9\,^{+2.1}_{-2.2}$ &$69.4$& $ 68.7^{+2.8}_{-2.6}$  &$67.4$   \\

$\sigma_8$   & $ 0.819 ^{+0.028}_{-0.030}$  &$83.9$ & $ 0.817\,\pm0.030$   &$0.819$& $ 0.816\,^{+0.024}_{-0.023}$ &$0.803$ \\

$\chi^2$   && $12979.9$   && $ 12984.3$   && $ 2466.7$  \\

\hline
\end{tabular}}

\end{table}

Regarding the inflationary parameters also in this $\Lambda$CDM+r+$N_{\rm eff}$ model we have a prediction for the tensor-to-scalar ratio $r$, which in our analysis is a derived parameter, different from zero: we find for Planck TT + lowP that $r=0.388\,^{+0.068}_{-0.061}$, and probably is this value not supported by the data to worsen the $\chi^2$ value. If we look at Figure~\ref{fignnu}, we can see that there exists a lower limit for $b>6.8\times10^7 GeV$ and $V_0=(46\pm10)\times 10^{-12} M_P^4$ for Planck TT+lowP. Finally, also when a dark radiation is included, we pass from the detection of a value of $\lambda_{hill}=0.309^{+0.079}_{-0.059}$ in Table~\ref{tablennup4} to just a lower limit $\lambda_{hill}>0.937$ in Table~\ref{tablennu}, and we obtain a stronger upper bound on $c_{hill}<0.00496$ for $\Lambda$CDM+r, now $c_{hill}<0.00127$ for $\Lambda$CDM+r+$N_{\rm eff}$, considering PlanckTT+lowP.

When considering PlanckTTTEEE+lowP, the BAO data or the ``tau055'' prior, we can see that the results are stable, so our conclusions are still valid.

Finally, in Table~\ref{tablew} there are the constraints for the $w$CDM+r scenario, by varying the equation of state of the dark energy, and using the Hilltop inflationary model with $p=6$, obtained in our analysis. From the comparison between the Table~\ref{tablew} and the constraints obtained in the same scenario for $p=4$ in Table~\ref{tablewp4}, we see a similar shift of the cosmological parameters than the previous cases. 
In particular, looking at PlanckTT+lowP we see a shift of $\Omega_ch^2$ and $\tau$ at about $1\sigma$ towards lower values, and of $H_0$ towards a higher one, while $\sigma_8$ is stable in this case. Also, in this modified Hilltop inflation with $p=6$, there is an indication for a dark energy equation of state $w<-1$ at about $3\sigma$, which disappears completely when adding the BAO dataset, restoring the Hubble tension. Again, we can notice that these shifts are not related to our modifications but to the Hilltop model with $p=6$ itself (Table~\ref{tableunmod_p6}). 

Regarding the inflationary parameters, we have still prediction for the tensor-to-scalar ratio $r$, which in our analysis is a derived parameter, that is $r=0.400\,^{+0.064}_{-0.061}$ for PlanckTT+lowP. Probably is this value not supported by the data to worsen the $\chi^2$ value of about $20$ if compared with the standard cosmological scenario for the same dataset. Moreover, we find a lower limit for $b$ at $b>7.1\times10^7 GeV$, stronger that the $p=4$ case, and $V_0=(48\pm10)\times 10^{-12} M_P^4$ for Planck TT+lowP. Finally, also when a constant dark energy equation of state is considered, we pass from the detection of a value of $\lambda_{hill}=0.311^{+0.088}_{-0.063}$ in Table~\ref{tablewp4} to just a lower limit $\lambda_{hill}>0.947$ in Table~\ref{tablew}, and we obtain a stronger bounds on $c_{hill}$, that is $c_{hill}=0.0029\,^{+0.0013}_{-0.0012}$ for $p=4$, while becomes $c_{hill}=0.00172\,^{+0.00094}_{-0.00096}$ for $p=6$, considering PlanckTT+lowP.

When analyzing PlanckTTTEEE+lowP or the ``tau055'' prior, we can deduce the same conclusions, while the addition of the BAO data changes only our conclusions on $w$ and $H_0$ as discussed before.

If we look at the Figures~\ref{clbf} and \ref{clbf_pol}, we can see the temperature and polarization power spectra obtained with the best fit of our modified Hilltop model with $p=6$ and the best fit of a minimal standard cosmological model $\Lambda$CDM+r, compared with Planck 2015 TT+lowP data. Our modified Hilltop model fits better the temperature large scales, improving the agreement with the data.

\begin{table}[]
\caption{$95 \% $ c.l. constraints on cosmological parameters in our baseline $w$CDM+r scenario from different combinations of datasets with a modified Hilltop inflation with $p=6$.}
\label{tablew}
\centering
\scalebox{0.9}{\begin{tabular}{lccccccc}
\hline
      {\bf Planck TT}    & &{\bf Best Fit}&&{\bf Best Fit}&&{\bf Best Fit} \\                     
 & {\bf + lowP}   &{\bf + lowP} &        {\bf + lowP + BAO} &{\bf + lowP + BAO}&{\bf + tau055}&{\bf + tau055} \\

$\Omega_{\textrm{b}}h^2$& $0.02239\,^{+0.00045}_{-0.00044} $&$0.02246$& $0.02227\,\pm0.00042$  &$0.02207$  & $0.02228\,^{+ 0.00043}_{-0.00042} $&$0.02230$   \\

$\Omega_{\textrm{c}}h^2$& $0.1156\,^{+0.0039}_{-0.0038}$&$0.1148$& $0.1165\pm 0.0035$  &$0.1169$  & $0.1177\,\pm0.0037 $  &$0.1182$ \\

$\tau$& $0.050\,\pm0.029$&$0.043$& $0.045^{+0.028}_{-0.031}$    &$0.044$& $0.055\,\pm 0.017$  &$0.052$ \\

$10^{12}V_0/M^4$& $48\,\pm10$&$44$& $48\,\pm10$    &$46$& $60\,\pm10$&$55$   \\

$log(b[GeV])$& $>7.85$&$8.6$& $>8.03$   &$14.3$ & $>8.08$ &$11.1$  \\

$10^{11}\lambda_{hill}$& $>0.947$&$0.988$& $>0.952$   &$0.999$ & $>0.932$  &$0.998$ \\

$c_{hill}$& $0.00172^{+0.00094}_{-0.00096}$&$0.0016$& $0.00193^{+0.00091}_{-0.00093}$   &$0.0022$ & $0.00288^{+0.00079}_{-0.00084}$ &$0.0026$  \\

$r$ &  $0.400\,^{+0.064}_{-0.061}$ &$0.38$&  $0.398\,^{+0.065}_{-0.064}$ &$0.39$ & $0.466\,^{+0.087}_{-0.077}$&$0.43$\\

$w$& $-1.61^{+0.40}_{-0.31}$&$-1.74$& $-0.96^{+0.12}_{-0.13}$   &$-1.00$ & $-1.56^{+0.53}_{-0.42}$   &$-1.68$\\

$H_0$ &      $>81$&$96.8$&      $ 67.2^{+3.0}_{-2.9}$&$68.2$ & $ 87 \,^{+10}_{-20}$   &$91.2$  \\

$\sigma_8$   & $ 0.97 ^{+0.09}_{-0.12}$ &$0.997$  & $ 0.779^{+0.045}_{-0.043}$   &$0.792$& $ 0.96^{+0.12}_{-0.15}$  &$0.994$\\

$\chi^2$   && $11289.9$  && $ 11302.8$   && $ 778.5$  \\
\hline

     {\bf Planck TTTEEE}    & & {\bf Best Fit} &&  {\bf Best Fit} &&  {\bf Best Fit} \\                     
 &  {\bf + lowP}   &  {\bf + lowP}  &        {\bf + lowP + BAO } &  {\bf + lowP + BAO} &  {\bf + tau055} &  {\bf + tau055} \\

$\Omega_{\textrm{b}}h^2$& $0.02230^{+0.00031}_{-0.00030} $&$0.02234$& $0.02224\,\pm0.00030  $   &$0.02215$ & $0.02226\,\pm0.00029 $  &$0.02234$ \\

$\Omega_{\textrm{c}}h^2$& $0.1176 \pm 0.0027$&$0.1173$& $0.1182\pm 0.0025$   &$0.1183$ & $0.1186 \,\pm0.0026$  &$0.1184$ \\

$\tau$& $0.043 \,\pm0.027$&$0.43$& $0.043\pm0.026$  &$0.047$  & $0.054\,\pm0.016$  &$0.054$ \\

$10^{12}V_0/M^4$& $49\,\pm10$&$47$& $50\,^{+10}_{-9}$   &$49$ & $59\,\pm10$ &$53$  \\

$log(b[GeV])$& $>7.95$&$17.3$& $>8.05$  &$14.4$  & $>7.88$  &$11.0$ \\

$10^{11}\lambda_{hill}$& $>0.960$&$0.991$& $>0.963$  &$0.988$  & $>0.951$ &$0.997$  \\

$c_{hill}$& $0.00215\,^{+0.00074}_{-0.00078}$&$0.0022$& $0.00224^{+0.00072}_{-0.00073}$   &$0.0021$ & $0.00245^{+0.00064}_{-0.00066}$ &$0.0022$  \\

$r$ &  $0.403\,^{+0.061}_{-0.058}$ &$0.39$&  $0.407\,^{+0.057}_{-0.055}$ &$0.40$ & $0.460\,^{+0.070}_{-0.066}$&$0.42$\\

$w$& $-1.66^{+0.42}_{-0.32}$&$-1.77$& $-1.01^{+0.11}_{-0.12}$   &$-1.05$ & $-1.60^{+0.51}_{-0.41}$  &$-1.59$ \\

$H_0$ &      $>81$&$95.1$&      $ 68.0\,^{+3.1}_{-2.8}$ &$69.1$& $ 88\,^{+10}_{-20}$   &$87.4$  \\

$\sigma_8$   & $ 0.98 ^{+0.09}_{-0.12}$  &$1.01$ & $ 0.797^{+0.037}_{-0.036}$  &$0.812$ & $ 0.97^{+0.11}_{-0.14}$ &$0.972$ \\

$\chi^2$   && $12973.7$  && $ 12986.9$   && $ 2459.9$  \\

\hline

\end{tabular}}

\end{table}

\section{Conclusions}
\label{sec:conclusions}

The quantum landscape multiverse describes the emergence of the universe from a wavefunction on the landscape before inflation, to a present-day classical universe. Other branches of the wavefunction, originating similarly to ours, are entangled with our universe. This quantum entanglement contributes as a second source a correction term in the gravitational potential of the universe, and it gives rise to modifications of the inflation potential and field evolution.

These modifications, first predicted in~\cite{tomolmh1,tomolmh2,tomolmh3,tomolmh4} and then~\cite{lmh} for concave potentials,  produce a series of anomalies, such as a suppressed $\sigma_8$, a giant void of size $200$ Mpc, suppressed spectrum at low multipoles, and so on. Previously, we checked the status of these predictions with Planck 2015 collaboration data for the exponential and Starobinsky type models of inflation in~\cite{eleonoraS,eleonoraE}. Here we complete our analysis of the status of the predictions against data with the investigation of a class of concave potential models, the Hilltop potentials.

We ran our analysis for the combined data sets, for the cases $p=4$ and $p=6$ of Hilltop models. Both these models allow a range of $b$ where the slow roll regime still holds, and all the predicted anomalies, including the giant void (cold spot)and the suppressed $\sigma_8$, are in very good agreement with data. By considering the quantum entanglement correction of the multiverse, we can place just a lower limit on the local 'SUSY-breaking' scale, respectively $b>8.7\times10^6$ GeV at $95 \%$ c.l. and $b>1.3\times10^8$ GeV at $95 \%$ c.l. from Planck TT+lowP, so the case with multiverse correction is statistically indistinguishable from the case with an unmodified inflation. 

Interestingly, the model of $p=6$ Hilltop inflation, goes beyond the agreement with the datasets for the spectrum and the confirmation of anomalies. This model also reduces the friction between the two major experiments on the value of the Hubble parameter: for $p=6$ the friction on the Hubble parameter disappears. Moreover, the $S_8$ values obtained is now perfectly consistent with the weak lensing experiments. However, this agreement is a characteristic of the Hilltop inflation and not of the modification due to the multiverse.

While we are excited that the anomalies predicted in this theory are in good standing with data independently of the chosen inflationary model, nevertheless we are certainly not claiming that the $p=6$ Hilltop model including the entanglement corrections from the quantum landscape multiverse, is the only allowed model of inflation. However, it is intriguing and encouraging that such an example where the anomalies are explained and the friction in the Hubble parameter and the $S_8$ value is removed, without introducing additional ingredients, does exist.

\section*{ACKNOWLEDGMENTS}
EDV acknowledges support from the European Research Council in the form of a Consolidator Grant with number 681431. LMH acknowledges support from the Bahnson funds.

\appendix
\section{}
\unskip
\begin{table}[H]
\caption{$68 \% $ c.l. constraints on cosmological parameters considering the minimal standard cosmological $\Lambda$CDM model and its extensions, for different combinations of datasets.}
\label{tablelcdm}
\centering
\scalebox{0.78}{\begin{tabular}{lccccccc}
\hline
          &{\bf Planck TT} & {\bf Planck TTTEEE}&{\bf Planck TT}& {\bf Planck TTTEEE}&{\bf Planck TT}& {\bf Planck TTTEEE}\\                     
         & {\bf + lowP}     &        {\bf + lowP}   & {\bf + lowP}       &  {\bf + lowP}& {\bf + lowP}       &  {\bf + lowP} \\

$\Omega_{\textrm{b}}h^2$&  $0.02224\,\pm0.00023 $& $0.02225\,\pm 0.00016$& $0.02230\,\pm 0.00037$ & $0.02220\,\pm 0.00024$ &  $0.02228\,\pm0.00023 $ & $0.02229\,\pm 0.00016$ \\

$\Omega_{\textrm{c}}h^2$&  $0.1195\,\pm0.0022 $& $0.1197\,\pm0.0014$ & $0.1205\,\pm 0.0041$ & $0.1191\,\pm 0.0031$ &  $0.1195\,\pm0.0022 $  & $0.1196\,\pm0.0015$\\

$\tau$&  $0.077\,\pm 0.019$& $0.078\,\pm 0.017$& $0.080\,\pm 0.022$ & $0.077\,\pm 0.018$  &  $0.076\,\pm 0.020$ & $0.075\,\pm 0.017$ \\

$log(10^{10}A_S)$&  $3.087\,\pm 0.036 $& $3.092\,\pm 0.033$& $3.096\,\pm 0.047$ & $3.088\,\pm0.038$ &  $3.085\,\pm 0.037 $& $3.085\,\pm 0.033$  \\

$n_S$&  $0.9666\,\pm0.0062$& $0.9652\,\pm0.0047$& $0.969\,\pm0.016$ & $0.9620\,\pm0.0097$  &  $0.9660\,\pm0.0061$& $0.9649\,\pm0.0048$  \\

$r$ & $<0.0472$&  $<0.0463$ &  $(0)$& $(0)$ &$(0)$ &$(0)$ \\

$N_{\rm eff}$ &   $ (3.046)$   &  $ (3.046)$ &  $ 3.13\,^{+0.30}_{-0.34}$  &  $ 2.99\,\pm0.20$&  $ (3.046)$ &  $ (3.046)$\\

$w$ &   $ (-1)$   &  $(-1)$ &  $ (-1)$  &  $ (-1)$ &  $ -1.54\,^{+0.20}_{-0.40}$&  $ -1.55\,^{+0.19}_{-0.38}$ \\

$H_0$ &       $ 67.42\,\pm0.99$   &  $ 67.31\,\pm 0.64$ &  $ 68.0\,^{+2.6}_{-3.0}$  &  $ 66.8\,\pm 1.6$ &       $>80.9$&       $>81.3$ \\

$\sigma_8$      & $ 0.828\,\pm 0.014$ &  $ 0.830\,\pm0.013$ &  $ 0.834\,^{+0.022}_{-0.025}$ &  $ 0.828\,\pm0.018$   & $ 0.98\,^{+0.11}_{-0.06}$ & $ 0.98\,^{+0.10}_{-0.06}$ \\

 \hline
         & {\bf Best Fit}  &       {\bf Best Fit}  &  {\bf Best Fit}      &   {\bf Best Fit} & {\bf Best Fit} &  {\bf Best Fit} \\

$\Omega_{\textrm{b}}h^2$&  $0.02224$& $0.02228$& $0.02224$ & $0.02217$ &  $0.02233 $ & $0.02230$ \\

$\Omega_{\textrm{c}}h^2$&  $0.1196 $& $0.1198$ & $0.1196$ & $0.1183$ &  $0.1191 $  & $0.1195$\\

$\tau$&  $0.080$& $0.083$& $0.078$ & $0.078$  &  $0.078$ & $0.0747$ \\

$log(10^{10}A_S)$&  $3.093 $& $3.101$& $3.089$ & $3.087$ &  $3.088 $& $3.082$  \\

$n_S$&  $0.9663$& $0.9659$& $0.969$ & $0.961$  &  $0.967$& $0.9654$  \\

$r$ & $0.0000$&  $0.0001$ &  $(0)$& $(0)$ &$(0)$ &$(0)$ \\

$N_{\rm eff}$ &   $ (3.046)$   &  $ (3.046)$ &  $ 3.04$  &  $ 2.94$&  $ (3.046)$ &  $ (3.046)$\\

$w$ &   $ (-1)$   &  $(-1)$ &  $ (-1)$  &  $ (-1)$ &  $ -1.94$&  $ -1.95$ \\

$H_0$ &       $ 67.38$   &  $ 67.32$ &  $ 67.34$  &  $ 66.52$ &       $100$&       $99.9$ \\

$\sigma_8$      & $ 0.831$ &  $ 0.834$ &  $ 0.829$ &  $ 0.826$   & $ 1.09$ & $ 1.09$ \\

$\chi^2$   & $11261.9$   & $ 12935.6$   & $ 11261.9$ & $ 12935.2$ & $ 11258.9$& $ 12932.3$\\
\hline

\end{tabular}}

\end{table}

\begin{table}[H]
\caption{$95 \% $ c.l. constraints on cosmological parameters considering the unmodified Hilltop inflationary model with $p=4$ for different combinations of datasets.}
\label{tableunmod_p4}
\centering
\scalebox{0.8}{\begin{tabular}{lccccccc}
\hline
          &{\bf Planck TT} & {\bf Planck TTTEEE}&{\bf Planck TT}& {\bf Planck TTTEEE}&{\bf Planck TT}& {\bf Planck TTTEEE}\\                     
         & {\bf + lowP}     &        {\bf + lowP}   & {\bf + lowP}       &  {\bf + lowP}& {\bf + lowP}       &  {\bf + lowP} \\

$\Omega_{\textrm{b}}h^2$&  $0.02224\,^{+0.00045}_{-0.00044} $& $0.02223\,^{+0.00031}_{-0.00030}$& $0.02234\,^{+0.00066}_{-0.00070}$ & $0.02220\,^{+0.00047}_{-0.00046}$ &  $0.02229\,\pm0.00045 $ & $0.02226\,\pm 0.00031$ \\

$\Omega_{\textrm{c}}h^2$&  $0.1196\,^{+0.0044}_{-0.0043} $& $0.1198\,\pm0.0029$ & $0.1208\,^{+0.0079}_{-0.0076}$ & $0.1192\,^{+0.0062}_{-0.0060}$ &  $0.1192\,^{+0.0044}_{-0.0043}$  & $0.1195\,\pm 0.0029$\\

$\tau$&  $0.077\,^{+0.038}_{-0.037}$& $0.079\,^{+0.032}_{-0.33}$& $0.081\,^{+0.041}_{-0.039}$ & $0.077\,^{+0.035}_{-0.034}$  &  $0.076\,^{+0.036}_{-0.037}$ & $0.074\,^{+0.034}_{-0.033}$ \\

$10^{12}V_0/M^4$& $<14.7$& $<13.7$    & $<14.5$ & $<13.4$   & $<14.9$ & $<16.8$ \\

$10^{11}\lambda_{hill}$& $0.307\,^{+0.076}_{-0.055}$& $0.311\,^{+0.069}_{-0.050}$    & $0.308\,^{+0.072}_{-0.052}$ & $0.309\,^{+0.069}_{-0.051}$ & $0.311\,^{+0.074}_{-0.056}$& $0.314\,^{+0.084}_{-0.059}$ \\

$c_{hill}$& $0.0031\,\pm 0.0013$& $0.00318\,\pm 0.00095$    & $<0.00499$ & $0.0034\,\pm 0.0018$& $0.0030\,\pm 0.0012$  & $0.0031\,\pm 0.0010$\\

$r$ & $<0.116$&  $<0.108$ &  $<0.116$& $<0.105$ &$<0.118$ &$<0.131$ \\

$N_{\rm eff}$ &   $ (3.046)$   &  $ (3.046)$ &  $ 3.17\,^{+0.55}_{-0.59}$  &  $ 3.00\,^{+0.41}_{-0.39}$&  $ (3.046)$ &  $ (3.046)$\\

$w$ &   $ (-1)$   &  $(-1)$ &  $ (-1)$  &  $ (-1)$ &  $ -1.54\,^{+0.59}_{-0.49}$&  $ -1.56\,^{+0.55}_{-0.46}$ \\

$H_0$ &       $ 67.4\,\pm1.9$   &  $ 67.3\,\pm1.3$ &  $ 68.4\,\pm 5.0$  &  $ 66.9\,\pm 3.1$ &       $>65$&       $>66$ \\

$\sigma_8$      & $ 0.828\,\pm 0.028$ &  $ 0.831\,\pm0.026$ &  $ 0.835\,^{+0.042}_{-0.041}$ &  $ 0.828\,^{+0.035}_{-0.034}$   & $ 0.98\,^{+0.14}_{-0.17}$ & $ 0.98\,^{+0.13}_{-0.16}$ \\
\hline
 
         & {\bf Best Fit}  &   {\bf Best Fit}  &  {\bf Best Fit}   &   {\bf Best Fit} & {\bf Best Fit} &  {\bf Best Fit} \\

$\Omega_{\textrm{b}}h^2$&  $0.02216 $& $0.02238$& $0.02210$ & $0.02224$ &  $0.02229$ & $0.02238$ \\

$\Omega_{\textrm{c}}h^2$&  $0.1222$& $0.1183$ & $0.1188$ & $0.1175$ &  $0.1193$  & $0.1185$\\

$\tau$&  $0.074$& $0.082$& $0.073$ & $0.103$  &  $0.088$ & $0.093$ \\

$10^{12}V_0/M^4$& $0.1$& $0.1$    & $2.8$ & $0.7$   & $3.0$ & $15.5$ \\

$10^{11}\lambda_{hill}$& $0.277$& $0.269$    & $0.296$ & $0.289$ & $0.299$& $0.389$ \\

$c_{hill}$& $0.0039$& $0.0030$    & $0.0039$ & $0.0034$& $0.0030$  & $0.0023$\\

$r$ & $0.001$&  $0.001$ &  $0.024$& $0.006$ &$0.025$ &$0.119$ \\

$N_{\rm eff}$ &   $ (3.046)$   &  $ (3.046)$ &  $ 2.91$  &  $ 3.00$&  $ (3.046)$ &  $ (3.046)$\\

$w$ &   $ (-1)$   &  $(-1)$ &  $ (-1)$  &  $ (-1)$ &  $ -0.89$&  $ -1.11$ \\

$H_0$ &       $ 66.3$   &  $ 67.9$ &  $ 66.0$  &  $ 66.9$ &       $64.3$&       $71.4$ \\

$\sigma_8$      & $ 0.838$ &  $ 0.826$ &  $ 0.822$ &  $ 0.843$   & $ 0.805$ & $ 0.870$ \\

$\chi^2$   & $11264.4$   & $ 12944.8$   & $ 11263.9$ & $ 12943.9$ & $ 11263.7$& $ 12942.6$\\
\hline

\end{tabular}}

\end{table}

\begin{table}[H]
\caption{$95 \% $ c.l. constraints on cosmological parameters considering the unmodified Hilltop inflationary model with $p=6$ for different combinations of datasets.}
\label{tableunmod_p6}
\centering
\scalebox{0.8}{\begin{tabular}{lccccccc}
\hline
          &{\bf Planck TT} & {\bf Planck TTTEEE}&{\bf Planck TT}& {\bf Planck TTTEEE}&{\bf Planck TT}& {\bf Planck TTTEEE}\\                     
         & {\bf + lowP}     &        {\bf + lowP}   & {\bf + lowP}       &  {\bf + lowP}& {\bf + lowP}       &  {\bf + lowP} \\

$\Omega_{\textrm{b}}h^2$&  $0.02233\,^{+0.00045}_{-0.00044} $& $0.02226\,\pm 0.00031$& $0.02264\,^{+0.00021}_{-0.00044}$ & $0.02258\,^{+0.00037}_{-0.00041}$ &  $0.02239\,\pm0.00045 $ & $0.02230\,^{+0.00031}_{-0.00030}$ \\

$\Omega_{\textrm{c}}h^2$&  $0.1158\,^{+0.0039}_{-0.0038} $& $0.1179\,\pm0.0027$ & $0.1234\,^{+0.0068}_{-0.0065}$ & $0.1232\,\pm 0.0057$ &  $0.1156\,^{+0.0039}_{-0.0038}$  & $0.1176\,^{+0.0027}_{-0.0028}$\\

$\tau$&  $0.049\,\pm 0.030$& $0.044\,\pm 0.027$& $0.055\,^{+0.026}_{-0.029}$ & $0.055\,^{+0.027}_{-0.029}$  &  $0.049\,\pm 0.029$ & $0.043\,^{+0.027}_{-0.026}$ \\

$10^{12}V_0/M^4$& $48\,\pm10$& $50\,\pm 10$    & $46\,\pm10$ & $48\,^{+10}_{-9}$   & $48\,\pm10$ & $49\,^{+10}_{-9}$ \\

$10^{11}\lambda_{hill}$& $>0.953$& $>0.962$    & $>0.941$ & $>0.957$ & $>0.948$& $>0.961$ \\

$c_{hill}$& $0.00177\,^{+0.00093}_{-0.00097}$& $0.00220^{+0.00075}_{-0.00077}$    & $<0.00127$ & $<0.00212$& $0.00174\,^{+0.00091}_{-0.00097}$  & $0.00216\,^{+0.00075}_{-0.00077}$\\

$r$ & $0.398\,^{+0.058}_{-0.057}$&  $0.405\,^{+0.059}_{-0.056}$ &  $0.386\,^{+0.061}_{-0.057}$& $0.399\,^{+0.057}_{-0.055}$ &$0.398\,^{+0.059}_{-0.057}$ &$0.402\,^{+0.060}_{-0.055}$ \\

$N_{\rm eff}$ &   $ (3.046)$   &  $ (3.046)$ &  $ 3.56\,^{+0.30}_{-0.32}$  &  $ 3.42\,^{+0.31}_{-0.35}$&  $ (3.046)$ &  $ (3.046)$\\

$w$ &   $ (-1)$   &  $(-1)$ &  $ (-1)$  &  $ (-1)$ &  $ -1.61\,^{+0.40}_{-0.32}$&  $ -1.66\,^{+0.41}_{-0.31}$ \\

$H_0$ &       $ 69.0\,\pm1.8$   &  $ 68.0\,^{+1.3}_{-1.2}$ &  $ 72.1\,^{+1.9}_{-2.2}$  &  $ 70.6\,^{+2.3}_{-2.5}$ &       $>81$&       $>81$ \\

$\sigma_8$      & $ 0.792\,^{+0.022}_{-0.021}$ &  $ 0.796\,\pm0.020$ &  $ 0.818\,^{+0.027}_{-0.028}$ &  $ 0.819\,^{+0.027}_{-0.030}$   & $ 0.97\,^{+0.09}_{-0.12}$ & $ 0.98\,^{+0.09}_{-0.11}$ \\

 \hline
         & {\bf Best Fit}  &   {\bf Best Fit}  &  {\bf Best Fit}   &   {\bf Best Fit} & {\bf Best Fit} &  {\bf Best Fit} \\

$\Omega_{\textrm{b}}h^2$&  $0.02223 $& $0.02223$& $0.02249$ & $0.02246$ &  $0.02236$ & $0.02223$ \\

$\Omega_{\textrm{c}}h^2$&  $0.1170$& $0.1179$ & $0.1183$ & $0.1224$ &  $0.1151$  & $0.1181$\\

$\tau$&  $0.041$& $0.034$& $0.031$ & $0.041$  &  $0.024$ & $0.033$ \\

$10^{12}V_0/M^4$& $50$& $54$    & $51$ & $48$   & $52$ & $54$ \\

$10^{11}\lambda_{hill}$& $0.965$& $0.933$    & $0.880$ & $0.957$ & $0.885$& $0.938$ \\

$c_{hill}$& $0.0020$& $0.0021$    & $0.0007$ & $0.0013$& $0.0015$  & $0.0022$\\

$r$ & $0.41$&  $0.45$ &  $0.45$& $0.41$ &$0.45$ &$0.44$ \\

$N_{\rm eff}$ &   $ (3.046)$   &  $ (3.046)$ &  $ 3.32$  &  $ 3.35$&  $ (3.046)$ &  $ (3.046)$\\

$w$ &   $ (-1)$   &  $(-1)$ &  $ (-1)$  &  $ (-1)$ &  $ -1.20$&  $ -1.15$ \\

$H_0$ &       $ 68.4$   &  $ 68.0$ &  $ 71.3$  &  $ 69.9$ &       $75.9$&       $72.7$ \\

$\sigma_8$      & $ 0.790$ &  $ 0.788$ &  $ 0.782$ &  $ 0.806$   & $ 0.826$ & $ 0.831$ \\

$\chi^2$   & $11296.6$   & $ 12979.9$   & $ 11288.7$ & $ 12979.0$ & $ 11288.3$& $ 12973.5$\\

\hline
\end{tabular}}

\end{table}

\end{document}